%% file: 4arch.tex
%%%%%%%%%%%%%%%%%%%%%%%%%%%%%%%%%%%%%%%%%%%%%%%%%%%%%%%%%%%%%%%%%%%%%%%%%
%                                                                       %
%        On FPL configurations with four sets of nested arches          %
%                                                                       %
%%%%%%%%%%%%%%%%%%%%%%%%%%%%%%%%%%%%%%%%%%%%%%%%%%%%%%%%%%%%%%%%%%%%%%%%%
% revised version of March 18

\input lanlmac.tex
\overfullrule=0pt
\input mssymb.tex

\writedefs

\font\eightrm=cmr8\font\eighti=cmmi8
\font\eightsy=cmsy8\font\eightit=cmti8
\font\eightsl=cmsl8\font\eighttt=cmtt8\font\eightbf=cmbx8
\font\sixrm=cmr6\font\sixi=cmmi6
\font\sixsy=cmsy6

\font\sixbf=cmbx6

\def\eightpoint{%
\textfont0=\eightrm \scriptfont0=\sixrm
\scriptscriptfont0=\fiverm \def\rm{\fam0\eightrm}%
\textfont1=\eighti \scriptfont1=\sixi
\scriptscriptfont1=\fivei \def\oldstyle{\fam1\eighti}%
\textfont2=\eightsy %\scriptfont2=\sixsy
\scriptscriptfont2=\fivesy
\textfont\itfam=\eightit \def\it{\fam\itfam\eightit}%
\textfont\slfam=\eightsl \def\sl{\fam\slfam\eightsl}%
\textfont\ttfam=\eighttt \def\tt{\fam\ttfam\eighttt}%
\textfont\bffam=\eightbf \scriptfont\bffam=\sixbf
\scriptscriptfont\bffam=\fivebf \def\bf{\fam\bffam\eightbf}%
\abovedisplayskip=9pt plus 2pt minus 6pt
\belowdisplayskip=\abovedisplayskip
\abovedisplayshortskip=0pt plus 2pt
\belowdisplayshortskip=5pt plus 2pt minus 3pt
\smallskipamount=2pt plus 1pt minus 1pt
\medskipamount=4pt plus 2pt minus 2pt
\bigskipamount=9pt plus 4pt minus 4pt
\setbox\strutbox=\hbox{\vrule height7pt depth2pt width 0pt} %
\normalbaselineskip=9pt \normalbaselines
\rm}
\def\petit{\vskip3mm\eightpoint
 \skewchar\eighti='177 \skewchar\sixi='177
\skewchar\eightsy='60 \skewchar\sixsy='60 }

%Definitions of Paul:
%
% redefine figures ... 
% works like standard \nfig except the figure is inserted in the
% middle of the text
\input epsf
\def\fig#1#2#3{
\xdef#1{\the\figno}
\writedef{#1\leftbracket \the\figno}
\nobreak
\par\begingroup\parindent=0pt\leftskip=1cm\rightskip=1cm\parindent=0pt
\baselineskip=11pt
\midinsert
\centerline{#3}
\vskip 12pt
%\petit{{\eightbf Fig.\ \the\figno:} {\eightrm #2}}\par
\petit{{\bf Fig.\ \the\figno:} {#2}}\par
\endinsert\endgroup\par
\goodbreak
\global\advance\figno by1
}
%
%Definitions of Philippe:
\def\file#1{#1}
\def\figbox#1#2{\epsfxsize=#1
\vcenter{\epsfbox{\file{#2}}}}
%\newcount\figno\figno=0
\def\figu#1#2#3{
\par\begingroup\parindent=0pt\leftskip=1cm\rightskip=1cm\parindent=0pt
\baselineskip=11pt
\global\advance\figno by 1
\midinsert
\epsfxsize=#3
\centerline{\epsfbox{#2}}
\vskip 12pt
{\bf Fig. \the\figno:} #1\par
\endinsert\endgroup\par
}
\def\figlabel#1{\xdef#1{\the\figno}}
\def\encadremath#1{\vbox{\hrule\hbox{\vrule\kern8pt\vbox{\kern8pt
\hbox{$\displaystyle #1$}\kern8pt}
\kern8pt\vrule}\hrule}}

%

%%%%%%%%%%%%%%%%%%%%%%%%%%%%%%%%%%%

\def\omit#1{}

\def\conf{configuration}
\def\CD{{\cal D}}\def\CO{{\cal O}}
\def\nind{\par\noindent}

\def\pre#1{({\tt
#1})}%use this to give preprint # in refs

\lref\Pro{J. Propp, {\sl The many faces of alternating-sign matrices},
preprint \pre{math.CO/0208125}.}
\lref\RS{A.V. Razumov and Yu.G. Stroganov, 
%{\sl Spin chains and combinatorics}, 
%\JPhA{34} (2001) 5335-5340, \condmat{0012141};  
{\sl Combinatorial nature
of ground state vector of O(1) loop model}, preprint \pre{math.CO/0104216}.}
\lref\MNdGB{S. Mitra, B. Nienhuis, J. de Gier and M.T. Batchelor,
{\sl Exact expressions for correlations in the ground state 
of the dense $O(1)$ loop model}, preprint \pre{cond-math/0401245}}
\lref\Wie{B. Wieland, {\sl  A large dihedral symmetry of the set of
alternating-sign matrices}, 
 {\it Electron. J. Combin.} {\bf 7} (2000) R37, 
\pre{math.CO/0006234}.}
\lref\dG{J. de Gier, {\sl Loops, matchings and alternating-sign
matrices},
% The art of number guessing: where combinatorics meets physics} 
preprint\pre{math.CO/0211285}.}
\lref\CaK{F. Caselli and C. Krattenthaler, 
{\sl Proof of two conjectures of Zuber on fully packed loop
configurations}, preprint\pre{math.CO/0312217}.}
\lref\PPJB{ P. Di Francesco, P. Zinn-Justin and J.-B. Zuber, 
{A  bijection between classes of Fully Packed Loops and plane
partitions},  preprint\pre{math.CO/0311220}.}
\lref\DMB{N. Destainville, R. Mosseri and F. Bailly,
{\sl Enumeration of octagonal random tilings by the Gessel-Viennot
method}, preprint\pre{math.CO/0302105}. }
\lref\dB{N.G. de Bruijn, {\sl Algebraic theory of Penrose's non-periodic
tilings of the plane}, {\it Kon. Neder. Akad. Wetensch. Proc. ser. A}
{\bf 43} (1981) 84; {\sl Dualization of multigrids}, {\it J. Phys. France,
Colloques C3}, {\bf 47} (1986) C3-9. }
\lref\LGV{B. Lindstr\"om, {\sl On the vector representations of
induced matroids}, {\it Bull. London Math. Soc.} {\bf 5} (1973)
85-90\semi
I. M. Gessel and X. Viennot, {\sl Binomial determinants, paths and
hook formulae}, {\it Adv. Math.} { \bf 58} (1985) 300-321. }
\lref\CK{C. Krattenthaler, {\sl Watermelon \conf s with wall
interaction: exact asymptotic results}, preprint.}
\lref\Kas{P.W. Kasteleyn, {\sl The statistics of dimers on a
lattice. I : The number of dimer arrangements on a quadratic lattice}, 
{\it Physica} {\bf 27} (1961) 1209-1225;
{\sl Dimer statistics and phase transitions},  
{J. Math. Phys.} {\bf 4} (1963) 287-293
\semi R. Kenyon, {\sl An introduction to the dimer model}, 
preprint\pre{math.CO/0310326}. }
\lref\Jock{W. Jockusch, {\sl Perfect matchings and perfect squares}, 
{\it J. Comb. Theory}, Ser A 67 (1994) 100-115.}
\lref\SalTo{N.C. Saldanha and C. Tomei, {\sl An overview of domino
and lozenge tilings}, {\it Resen. Inst. Mat. Estat. Univ. Sao Paulo
}{\bf 2}, No.2, (1995) 239-252, \pre{math.CO/9801111}.}
%
% 
% new refs
\lref\CEKZ{M. Ciucu, T. Eisenk\"obl, C. Krattenthaler and D. Zare,
{\sl Enumeration of lozenge tilings of hexagons with a central
triangular hole}, {J. Combin. Theory Ser. A} {\bf 95} (2001) 251-334;}
\lref\MNosc{S. Mitra and B. Nienhuis, {\sl 
Osculating random walks on cylinders}, {\it in}
{\it Discrete random walks}, 
DRW'03, C. Banderier and
C. Krattenthaler edrs, Discrrete Mathematics and Computer Science
Proceedings AC (2003) 259-264, \pre{math-ph/0312036} . } 
\lref\MNnew{S. Mitra and B. Nienhuis, {\sl Exact conjectured
expressions for correlations in the dense O(1) loop model on cylinders 
}, to  appear.}

%%%%%%%%%%%%%%%%%%%%%%%%%%%%%%%%%%%%%%%%%%%%%%%%%%%%%%%%%%%%%%%%%%%%%
%

\Title{SPhT-T04/030}
{\vbox{
\centerline{On FPL configurations with four sets of nested arches}
}}
\bigskip
\centerline{P.~Di~Francesco and J.-B. Zuber} 
\omit{\footnote{${}^\#$}
{Service de Physique Th\'eorique de Saclay, 
CEA/DSM/SPhT, URA 2306 du CNRS, 
C.E.A.-Saclay, F-91191 Gif sur Yvette Cedex, France}}
\medskip
\centerline{\it  Service de Physique Th\'eorique de Saclay,}
\centerline{\it CEA/DSM/SPhT, URA 2306 du CNRS,}
\centerline{\it F-91191 Gif sur Yvette Cedex, France}

\bigskip\noindent
The problem of counting the number of Fully Packed Loop (FPL)
configurations with four sets of $a,b,c,d$ nested arches is addressed. It is
shown that it may be expressed as the problem of enumeration of
tilings of a domain of the triangular lattice with a conic singularity. 
After reexpression in terms of  non-intersecting lines, 
the Lindstr\"om-Gessel-Viennot theorem leads to a formula 
as a sum of determinants. This is made quite explicit 
when $\min(a,b,c,d)=1$ or 2. We also find a compact determinant
formula which generates the numbers of configurations with $b=d$.


%\draft
\Date{03/2004}

%
%%%%%%%%%%%%%%%%%%%%%%%%%%%%%%%%%%%%%%%%%%%%%%%%%%%%%%%%%%%%%%%%%%%%%
%

Given a square grid of side $n$, Fully Packed Loops (FPL)
are sets of paths which visit once and only once each of the $n^2$
sites of the grid and exit through every second of the $4n$ 
external edges. FPL of a given size fall into connectivity classes, 
or {\it link patterns}, of \conf s with a definite set of 
connectivities between their external edges. The problem of
enumerating FPL of a given link pattern is a challenging problem
for the combinatorialist, related to
alternating sign matrices and other problems of current 
interest (see \refs{\Pro,\dG} for reviews).
It is also of relevance in statistical mechanics, as it is related 
by the Razumov-Stroganov conjecture \RS\ to the $O(1)$-loop model of
percolation, see \MNdGB\ for references.

This paper, which is  a continuation of \PPJB,  is devoted to a 
study of FPL configurations with four sets of nested arches. We
shall assume the reader to have some familiarity with the ideas and
techniques developed in \PPJB\ for the case of three sets
of nested arches. In particular, with the notion that the boundary
conditions force a certain number of edges to be occupied or
empty (``fixed edges''), see also \dG\ for a precursor of this idea 
and \CaK\  for a recent application to other types of FPL
configurations.

Our aim is not only to get formulas as explicit as
possible for the numbers of these FPL configurations, but also --and
mainly-- to see to what extent this problem is equivalent to the
counting
of tilings of certain domains of the triangular lattice, or in a dual
picture, to that of dimer \conf s on a certain graph. 

% Fig 1
{\fig\generic{The link pattern of
FPL \conf s with four sets of nested arches.}{\epsfbox{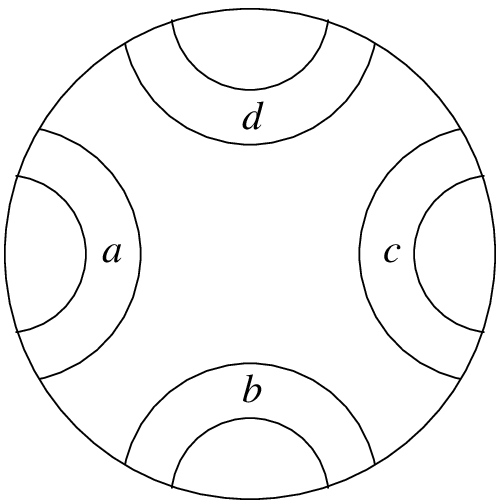}}}

We shall consider FPL \conf s with four sets of $a$, $b$, $c$ and $d$
nested arches and denote $A_n(a,b,c,d)$ their number, $n=a+b+c+d$
being the total number of arches, see Fig. 1. 
We recall Wieland's theorem \Wie, which asserts that $A_n(a,b,c,d)$
depends only on the orbit of the FPL link pattern under
the action of the dihedral group $D_{2n}$. Contrary to the simple
case of 3 sets of arches \PPJB, but like in the more complicated case 
treated in \CaK, we use this theorem to pick a particularly 
suitable representative of the orbit. 

% Fig 2
{\fig\generic{The generic situation where
$a>c$, $b>d$.  (Here $a=13$, $b=10$, $c=8$ and $d=1$). 
(i) the fixed occupied  edges in red or blue, the unfixed ones in
dotted lines; (ii) the octagonal domino grid $\CO$.}{\epsfbox{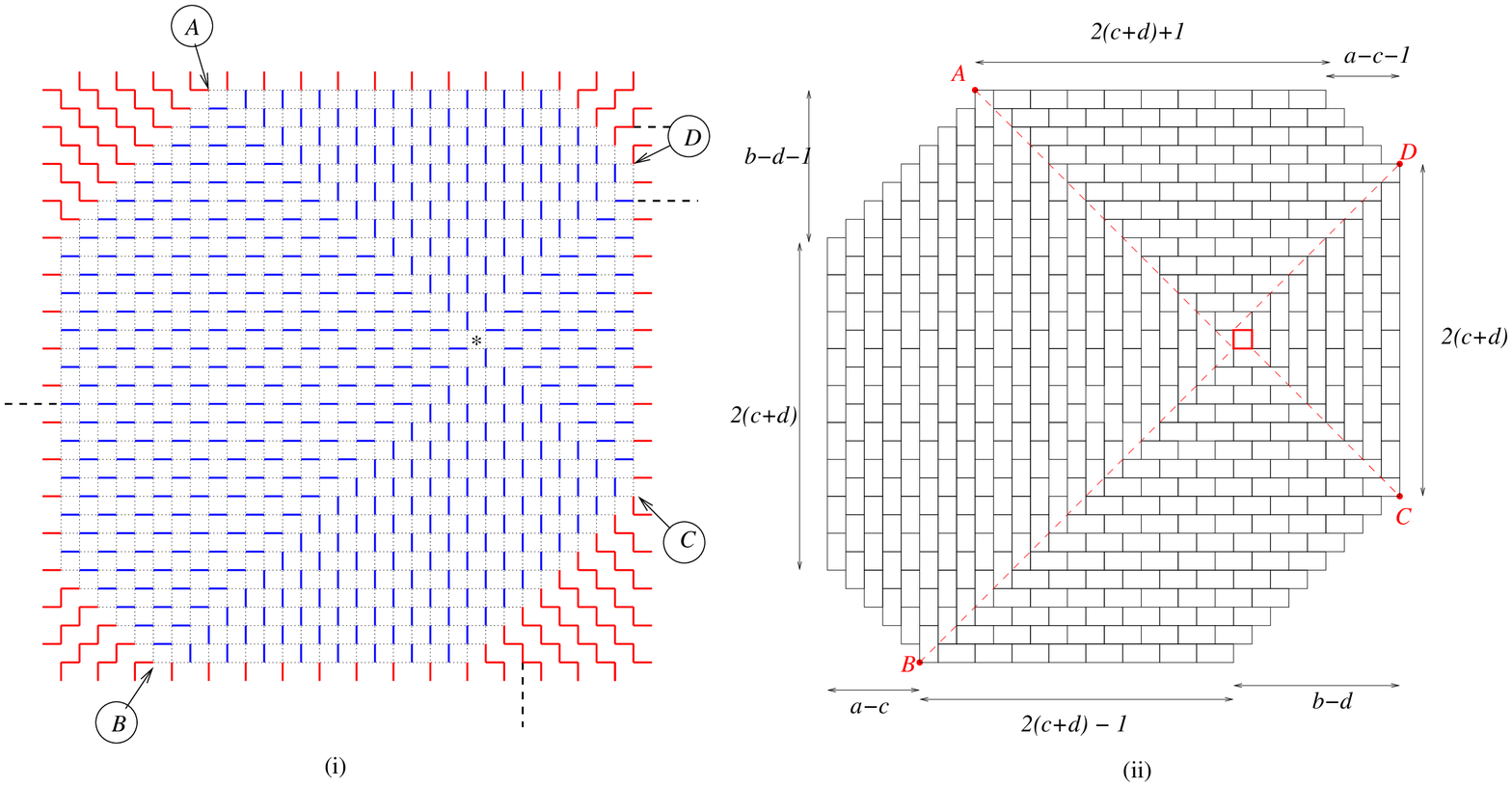}}}

\newsec{From FPL to dimers to non intersecting lines}
\subsec{The octagonal domain of unfixed edges} 
\noindent
Using the reflective and cyclic symmetries of the problem, 
$A_n(a,b,c,d)=A_n(a,d,c,b)=A_n(b,c,d,a)$, we may always assume that
$a\ge c$ and $b\ge d$. Let $A,B,C,D$ denote the centers 
of the sets of nested arches, 
we assume they are in anticlockwise order. Because 
 $2(a+b)\ge n$, $A$ and $B$ belong to different sides of the
square. Then  one uses the same procedure as in 
\refs{\dG,\PPJB} to fix edges in $45^{\rm o}$-cones of vertices $A,B,C,D$:
outside these cones  the occupied edges either form staircases (red edges on 
Fig.\generic\ (i)),  or every
second (blue) edge parallel to the external
ones is occupied: we refer the reader to \PPJB\ for a discussion
of the procedure. 
The unfixed edges then live either on the sides of rectangular 
$2\times 1$ tiles, also called dominos, the inner sides of which are
occupied, or inside a connected domain made of adjacent elementary
squares. Because the latter squares appear as defects (``disclinations''
in the language of cristallography)
in the tiling reinterpretation of the FPL, we try to make this 
domain as small as possible. The choice of a Wieland rotation
which brings the diagonal lines emanating from $A$ and $C$ 
and those from $B$ and $D$ almost colinear, see Fig. \generic\ (ii), 
reduces this defect zone to a single elementary square (indicated by a 
star in Fig. \generic\ (i) and drawn in red in Fig. \generic\ (ii)). Thus

\medskip
\noindent{\bf Proposition 1.} {\sl If one chooses Wieland's rotation such
that
the centers of arches are as depicted in Fig. \generic, 
every site of the lattice belongs to at least one fixed edge, and 
the unfixed edges form an octagonal pattern $\CO$  made of 
dominos surrounding a single elementary square.}
\smallskip

The locations of the points $A,B,C,D$ and of the
small  square are determined by the data given in Fig. \generic.
We refer to this pattern of unfixed edges as the 
{\it domino grid} $\CO$, to the sites 
which belong to a single fixed edge as {\it active}, and
to the elementary square as the {\it central square}.

The figure depicts the generic situation when both differences 
$a-c$ and $b-d$ are greater than 0. This includes the cases where
one or/and  the other equals 1, and where the octagon loses some side(s)
and acquires right angles. 
% Fig 3
{
 \fig\limiting{Limiting cases where $a-c$ and/or $b-d$ equals 0 or 1.
}{\epsfbox{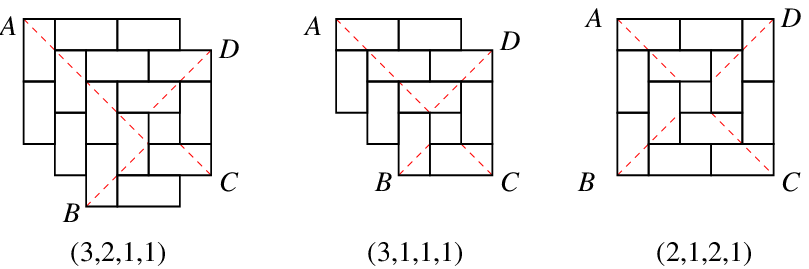}}}
In the case $a-c=0, b-d>0$ (or vice versa),
the side of length $a-c-1$ shrinks to naught, while the
adjacent sides  are reduced by one unit, hence have lengths
$2(c+d)-1$ and $2(c+d)$. One is left with 
a hexagonal domain with two right angles, see Fig. \limiting.
If both $a=c$ and $b=d$, one gets a rectangle. 

\smallskip

% Fig 4 
{
\fig\fullemptyd{The dimers of the empty and the  full 
\conf s, for $(a,b,c,d)=(9,14,4,5)$. 
The broken black lines show how to define the limits of the various domains,
while the pink ones give the limits of the sets of nested arches. 
The light blue dots will be explained below in sect. 1.4.3.}
{\epsfbox{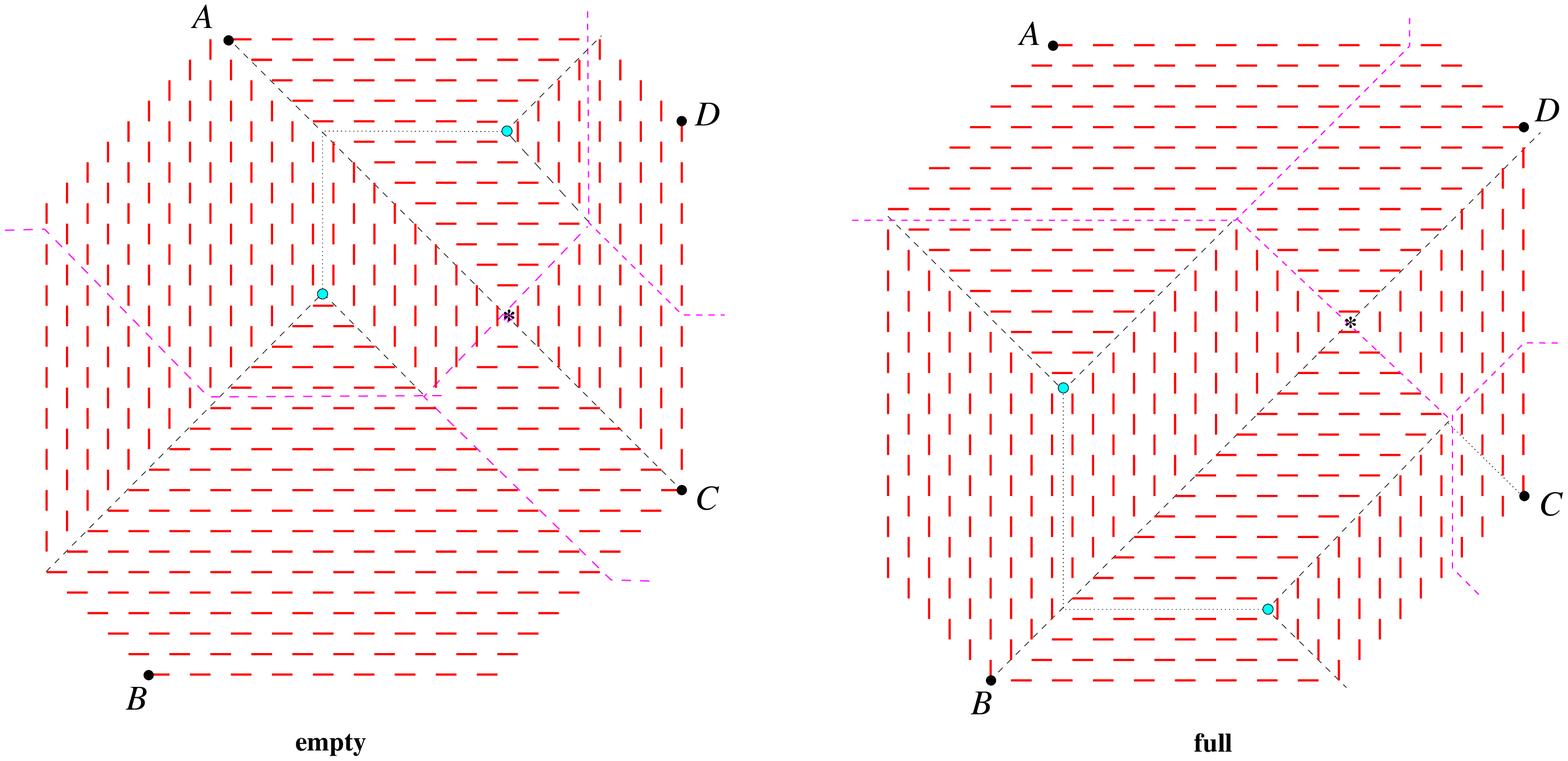}}}
Given this set of fixed edges, an FPL \conf\ is determined by an 
appropriate choice of {\it dimers}, i.e.,  of pairings, 
between the active sites, realizing the desired connectivity pattern. 
In particular, there are two special configurations, that we call the
``empty''  and the ``full'' ones. 
These \conf s are obtained by dividing the domain $\CO$ into subdomains 
and choosing the dimers as indicated in Fig. \fullemptyd.

Thus 
to any FPL \conf\ corresponds a dimer \conf. This correspondence between 
the set of FPL \conf s of type $(a,b,c,d)$ and the set of dimer
\conf s on $\CO$ cannot, however, be one-to-one.
Indeed it is clear from our discussion that the domino grid is common to
all FPL types $(a+p,b-p,c+p,d-p)$ for all $p$, 
$-c \le p \le d$.  This is obvious in Fig. \generic\ where the effect of
$(a,b,c,d) \to (a\pm 1, b\mp 1, c\pm 1, d \mp 1)$ is just to
shift the boundaries between the different sets of nested arches 
(the broken black
lines in Fig.\generic\ (i)),  while preserving the points $A,B,C,D$, the
shape of the octagon, the domino grid and the location of the central %small 
square. 

To distinguish dimer \conf s pertaining to different $p$'s, 
we now introduce a new feature, made of  non intersecting lines.

%%%%%%%%%%%%%%%%%%%%%%%%%%%%%%%%%%%%%%%%%%%%%%%%%%%%%%%%%%%%%%%%%%%%%%%

\subsec{Tilings and de Bruijn lines}

% Fig 5
{
\fig\triangul{Triangulating the octagonal grid. Here (i) $a=2,b=c=d=1$
and (ii), (iii) 
$a=9, b=14,c=4,d=5$. }{\epsfbox{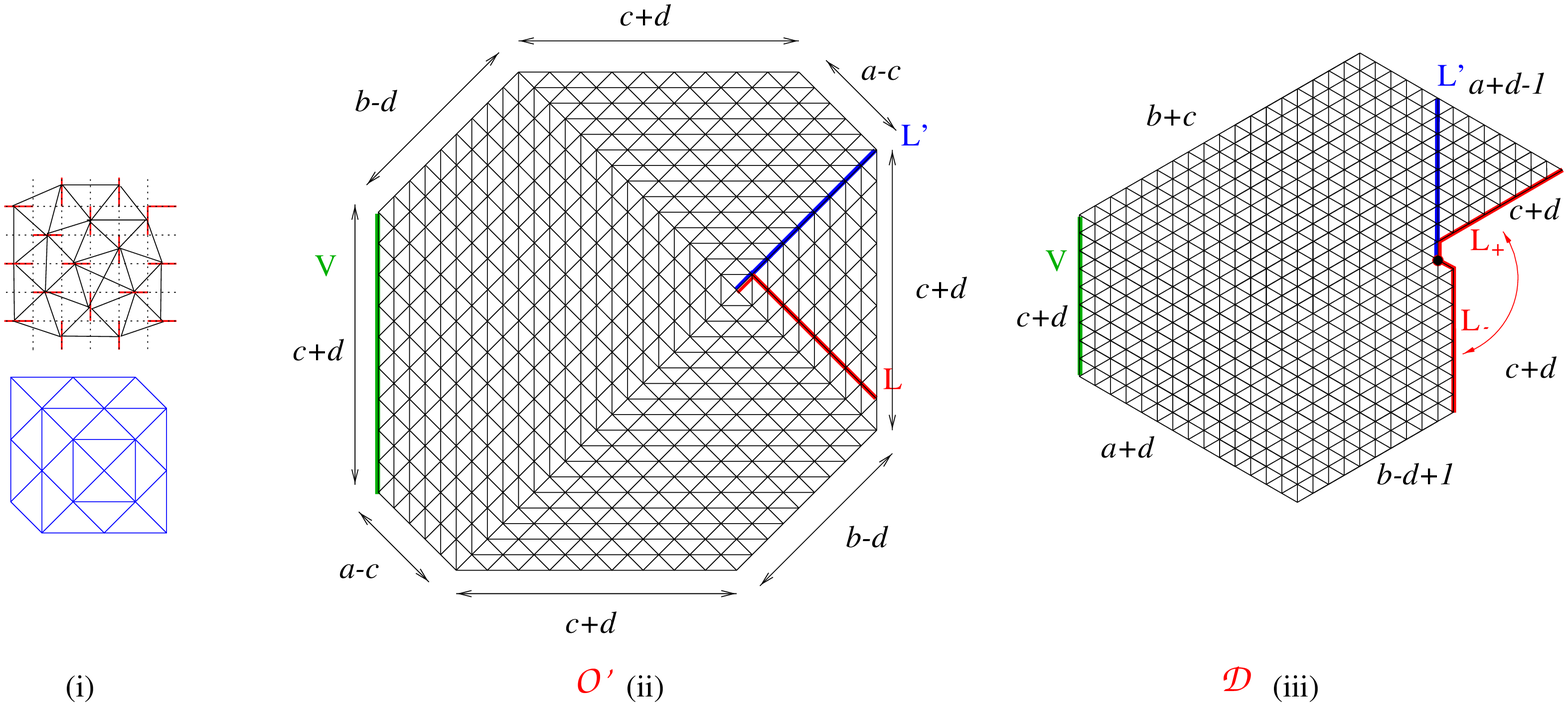}}}

At this stage we find it useful to introduce the dual picture, where 
one constructs a triangle around each active site. The central %small 
square
is also triangulated by four triangles. One gets an octagonal picture $\CO'$
which is depicted in Fig. \triangul, in (i) for the simple FPL $(2,1,1,1)$
and in (ii) for a more generic one.

A final transformation consists in cutting this octagon along a line
L starting from the central square, deforming the grid into a
domain $\CD$ of the regular triangular lattice and identifying the two 
sides L$_\pm$ of the cut. 
Figure \triangul\ (iii) shows one particular way 
to do this. For future use, we also draw a segment 
L' which joins the center of the central square to 
the ``East-North-East'' corner of the octagon. 
(Note that 
the segments L and L' do not coincide with the lines used in the construction 
of Fig. \generic, although they are parallel and close to them.
{Below, we shall slightly modify them,  
in a way depending on the tiling at hand, so as to
prevent them from intersecting the tiles.})

% Fig 6
{
\fig\tiles{The different types of tiles: (i) on the octagonal grid,
(ii)
after its deformation to equilateral triangles.
}{\epsfbox{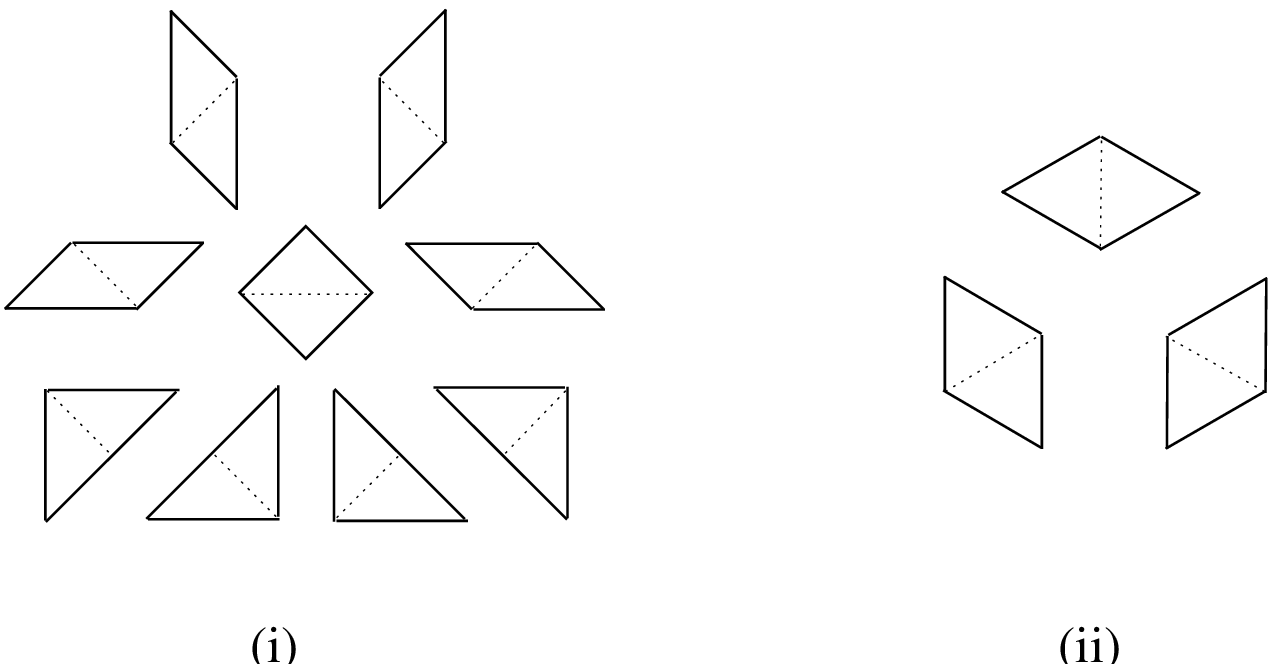}}}

Any FPL \conf\ yields a 
pairing between triangles sharing an edge, hence a tiling of the
octagon by means of the tiles depicted in Fig. \tiles.
Just as in the previous section, the tilings of the domain $\CD$ 
comprise all cases $(a+p,b-p,c+p,d-p)$, for $p$ running from 
$-c$ to $d$.

For a particular $(a,b,c,d)$, we have to find a refined 
characterization of  the tiling \conf s. We do this by means of an
alternative representation by systems of non intersecting lines, 
also known as de Bruijn curves \refs{\dB,\DMB}.

% Fig 7
{
\fig\hexagon{The alternative descriptions of a plane partition,
(i) as a set of dimers on the honeycomb lattice, 
(ii) as a tiling or stack of cubes, and 
(iii) as a system of non-intersecting lines of either of three colours.
}{\epsfbox{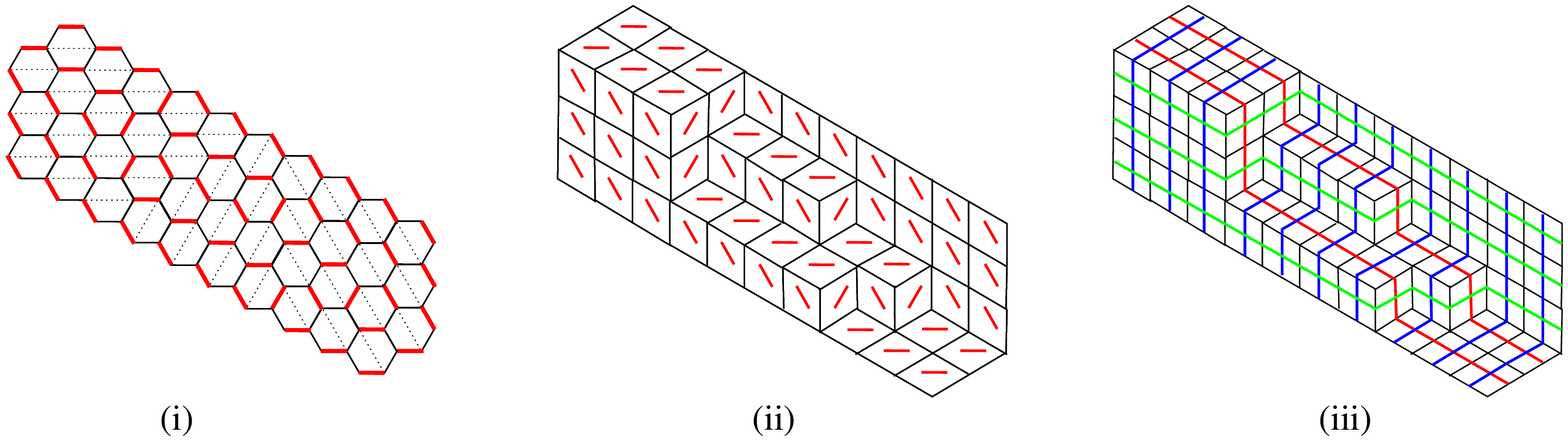}}}

Recall that in the simple case of the tiling of a hexagon of size
$a\times b\times c$, or equivalently of the plane partitions in a box
of that size, there is a one-to-one correspondence between 
(i) \conf s of dimers on   a domain of the honeycomb lattice, 
(ii) \conf s of tiles (in the tiling problem of the hexagon)  or of
elementary cubes (in the plane partition picture), 
and (iii) any of the three 
families of non intersecting lines joining a pair of opposite sides of the
hexagon, see Fig. \hexagon. Each family 
describes a collection of strips of 
tiles, which pairwise  share edges parallel to a given direction.

\def\ta{{\bf a}}\def\tb{{\bf b}}\def\tc{{\bf c}}\def\td{{\bf d}}

%%%%%%%%%%%%%%%%%%%%%%%%%%%%%%%%%%%%%%%%%%%%%%%%%%%%%%%%%%%%%%%%%

\subsec{The $\tc$ and $\td$ lines: definition and properties}

% Fig 8
{\fig\tilfulempt{Tiling of the empty and full \conf s. }{\epsfbox{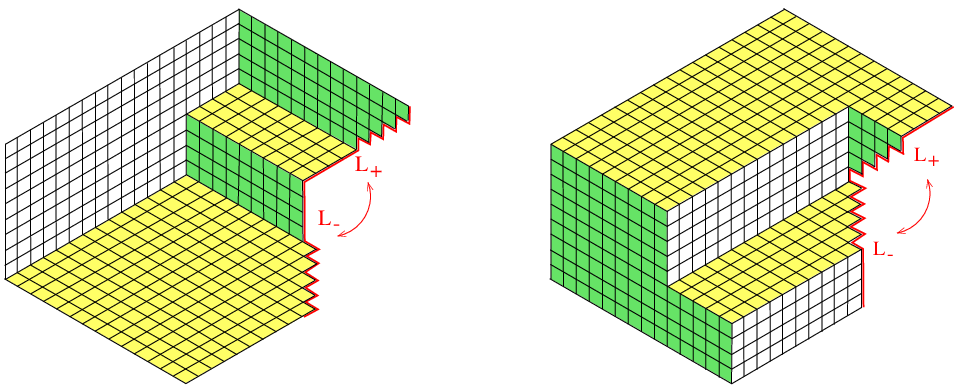}}}

% Fig 9
{\fig\tilfulemptl{The de Bruijn lines of the empty and full \conf
s. The segment L' appears in red, slightly deformed 
so as to follow tile edges. }{\epsfbox{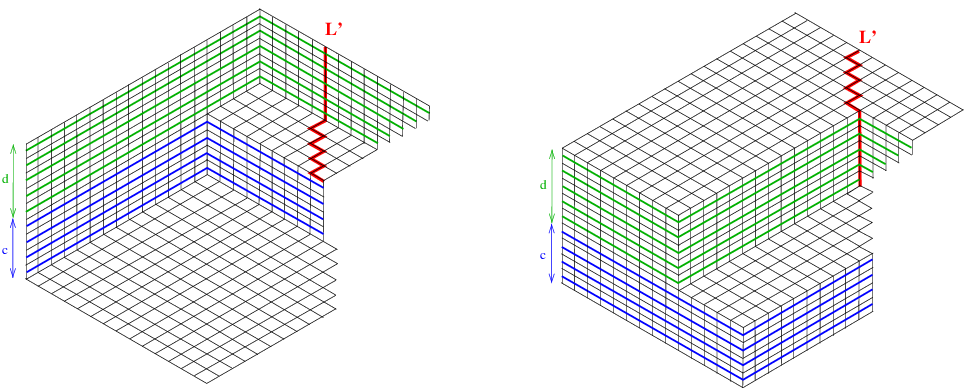}}}

% Fig 10
%\figu{}{setting.eps}{12.cm}\figlabel\setting
{\fig\setting{The $\tc$ (blue) and $\td$
(green) lines of a
FPL \conf\ of type $(9,14,4,5)$. On the right figure, they
 have been continued 
across the cut as broken lines. }{\epsfbox{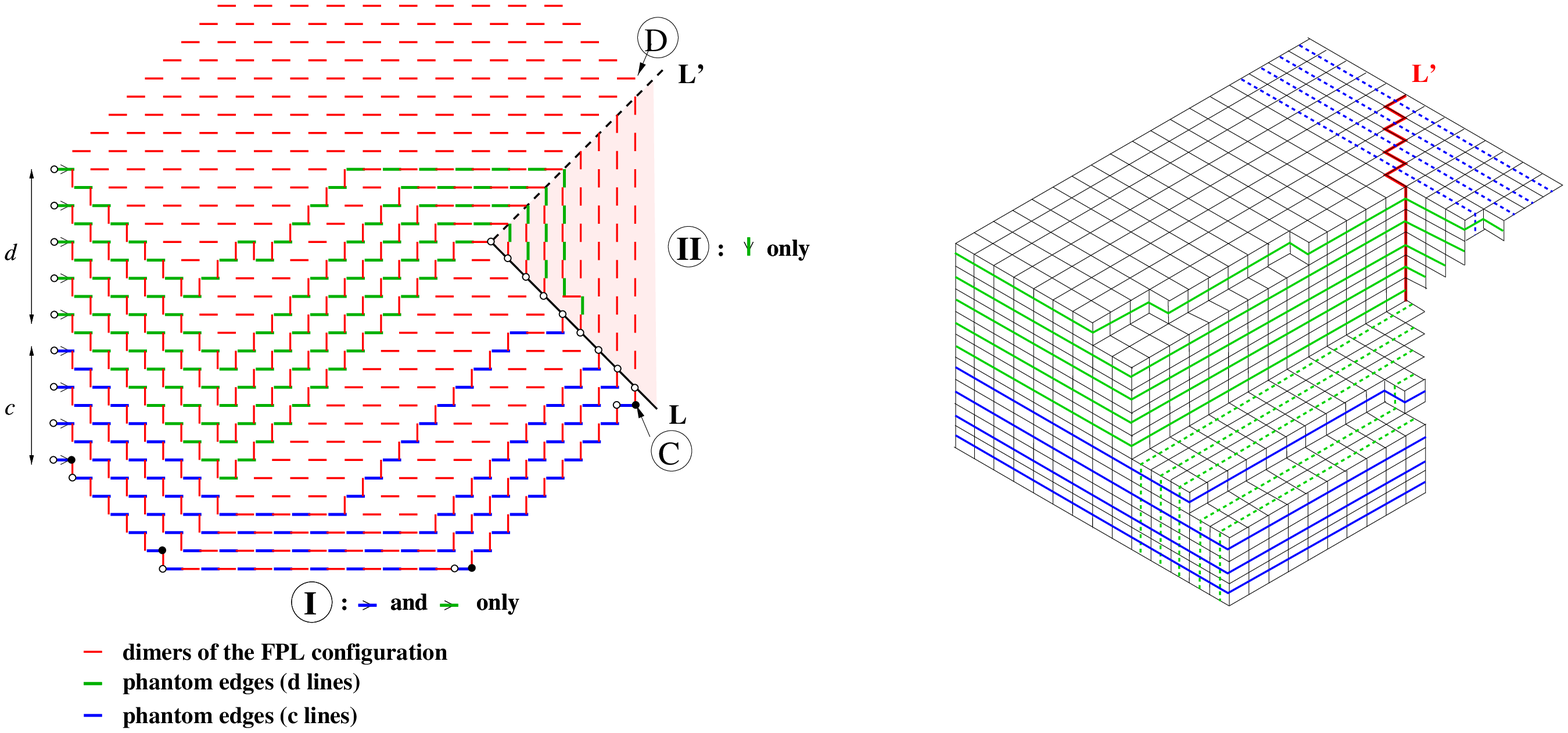}}}

\noindent Given the tiling associated with a certain FPL \conf\ of
type $(a,b,c,d)$, consider the $c+d$ de Bruijn lines which start 
from the vertical left side V of the domain $\CO'$ of Fig. 
\triangul~(ii) or $\CD$ of Fig.  \triangul~(iii). 
As noticed above, the tiling is made of strips of tiles, pairwise
sharing edges parallel to a given direction, and the de Bruijn lines
join the middles of these edges:  see for 
example the tilings and these lines for the empty and full \conf s
in Fig. \tilfulempt\ and \tilfulemptl. 
These lines are non-intersecting, and we call  $\tc$ lines the first
$c$ starting from the bottom of the left vertical interval V,
and $\td$ lines, those (in number $d$) which 
start from the upper part of that interval. 
We continue these 
lines across the domain until they reach the line L or exit through the 
boundary of $\CO'$, whichever occurs first. 
In fact, we claim (and will prove below) 
that they all reach the line L first, and more precisely, that 
the $\tc$ lines will reach it from below, while the $\td$ ones 
will do it from above.
 On the domain $\CD$, the $\tc$ lines reach the lower cut L$_-$ while
the $\td$ lines reach the upper one L$_+$. 
We now state the main result of this paper:

\nind{\bf Theorem 1.} 
{\sl There is a bijection between FPL \conf s of type 
$(a,b,c,d)$ and families of $\tc$ and $\td$ non intersecting lines on $\CD$, 
where the $\tc$ lines go from V to L$_-$, 
and the $\td$ lines go from V to L$_+$. The
sets of points where the $\tc$ lines and the $\td$ lines reach L
are disjoint.}

This theorem has been stated for the lines on $\CD$ but can of course
be rephrased on $\CO'$. 
Because the endpoints on L of  the $\td$ lines are disjoint 
from those of the $\tc$, we allow a slight (configuration-dependent) 
redefinition of the 
lines L$_\pm$ on the domain $\CD$ so as to make them lie along  tile edges, 
see Fig. \tilfulempt.

\medskip To prove this theorem, we establish a certain number of 
properties of the $\tc$ and $\td$ lines, going back and forth between the
two pictures, on the octagonal grid $\CO$ on the one hand, on the cut 
domain $\CD$  one on the other. The properties of the lines are 
indeed easier to establish in the tiling version, but to 
discuss their interplay
with the FPL paths, it is essential to return to  $\CO$. 

The first step is thus to translate the 
construction of the $\tc$ and $\td$ lines back to the octagonal grid
$\CO$. By a slight abuse of notation, we still denote them by 
$\tc$, $\td$ on $\CO$.
Note that the (active) sites of $\CO$ are  bicolorable. By convention, 
we assign the color $\bullet$ to point $C$ 
and to all sites distant from it by an even number of lattice steps, 
and $\circ$ to the others. (For the sake of clarity, only a few
sites have been colored in Fig. \setting.)
When drawn on $\CO$, the two segments L and L' start from the
upper right corner of the central square and 
pass by the external endpoints of the two empty edges entering 
the FPL grid at the points $C$ and $D$, respectively
\foot{This last assertion has to be slightly amended in the
 case where $a=c, b=d$ and the four points $A,B,C,D$ 
lie at the corners of the octagon, which has degenerated into a square
(see Fig. \limiting).}. 
 These lines divide the octagonal domain $\CO$ 
into two regions I and II, see Fig. \setting. 
Note that, by the construction of the 
subsection 1.1, region II contains all the horizontal fixed edges 
of the right part of $\CO$ and segments L and L' pass through their leftmost
endpoints.
Now, for a given FPL configuration of  type $(a,b,c,d)$, associated with a 
certain choice of dimers on the domino grid $\CO$, 
we call {\it vacant} the yet unused edges of
$\CO$. The  $\tc$ and $\td$ lines start from the $c+d$
vacant horizontal external edges 
bordering the left vertical side of the
domino grid and are oriented inwards the grid. 
They visit alternatingly  a vacant 
edge and a  dimer of the FPL configuration, with the rule that the chosen 
vacant edges (called {\it phantom}) are all horizontal 
(and oriented from left to right) in 
region I and vertical (and oriented from top to bottom) in region II. 
This rule is just the transcription of the definition of the
$\tc$ and $\td$ lines on $\CD$, namely it reflects the  pairing of adjacent
tiles which share a vertical edge.

In the same way as on $\CD$, the first $c$ lines from the bottom 
are called $\tc$ lines, the next
$d$ are called $\td$ lines, and we interrupt them as soon as they touch 
either the segment L or one of the boundaries of the grid. 
Note that the fact that the lines may be constructed without
encountering any obstruction or that they are non-intersecting, 
which is not obvious from the standpoint
of $\CO$, follows from their construction on the domain $\CD$ as 
conventional de Bruijn lines.  This is summarized in 

\nind 
{\bf Lemma 1. }{\sl  On $\CO$, the $\tc$ and $\td$ lines are
non intersecting; each of them is described 
by an alternate sequence of dimers and of ``phantom'' edges.}\nind
{\sl In region I the dimers
visited by the $\tc$ and $\td$ lines are of the three types}
$\!\!\!\!\!\!\figbox{24mm}{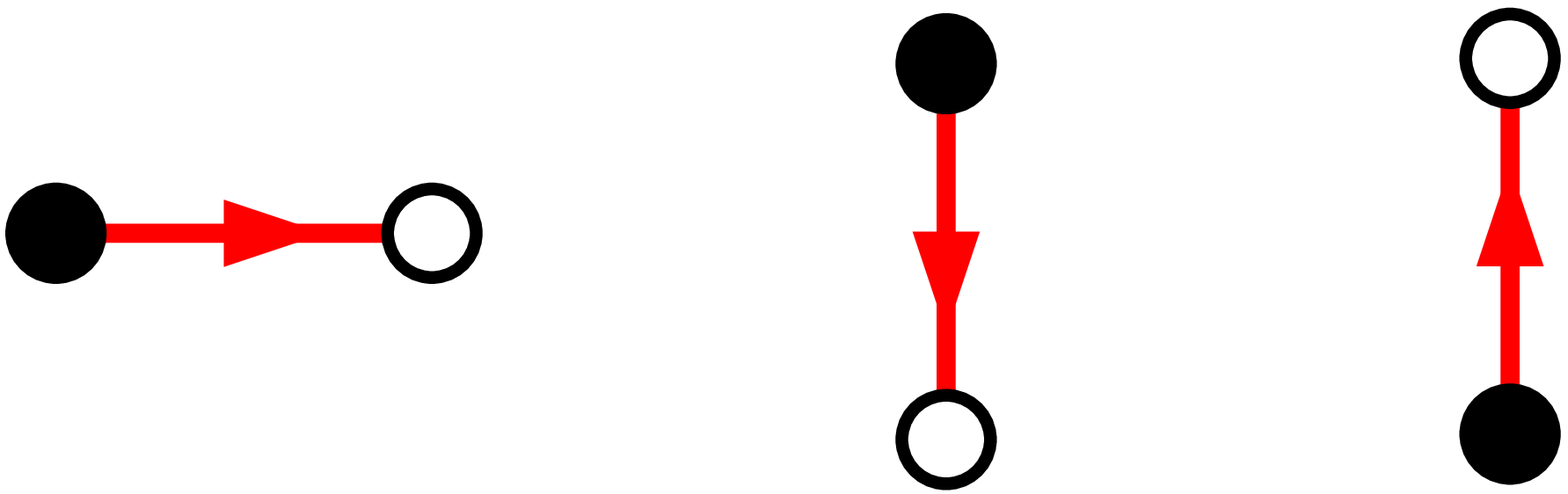} $ 
{\sl while\ in\ region\ II,\ they\ are:}
$\!\!\!\!\!\!{ \figbox{28mm}{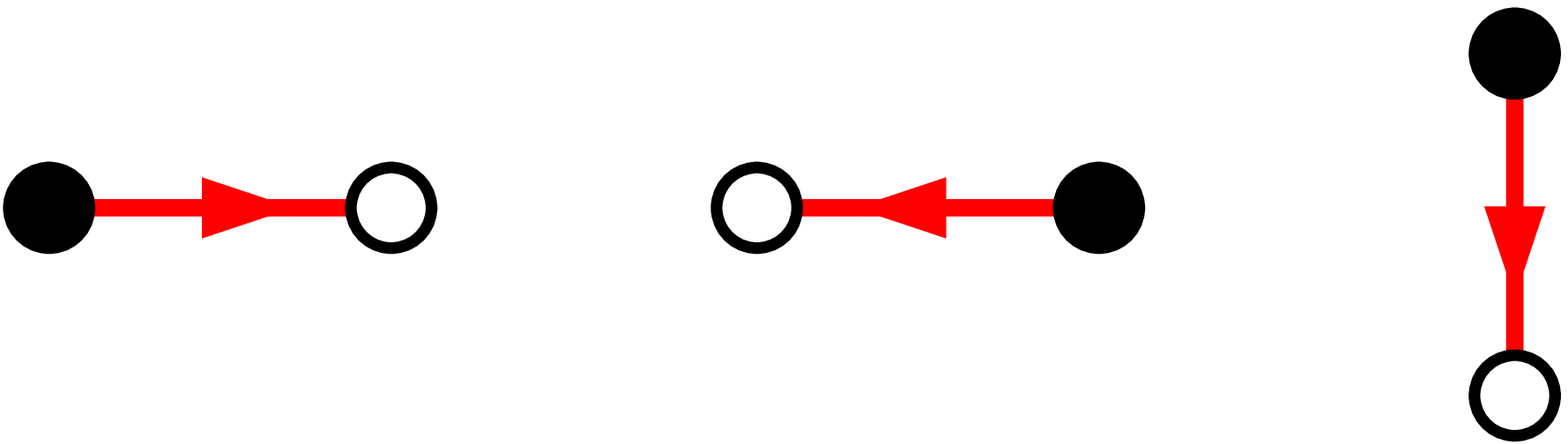} }$

Another property of $\tc$ and $\td$ lines
which is clear from their definition on $\CD$ 
is the fact that between two of them, or below the lowest one, or 
above the highest one, the tiling is frozen and uses only 
horizontal rhombi. This is easily established, 
starting from the leftmost corner of any domain of the triangular
lattice lying between two successive lines, or between the 
upper or lower one and the boundary, and proceeding iteratively.
This also means that the data of $\tc$
and $\td$ lines are sufficient to characterize the tiling entirely.
When restated on $\CO$, this property means that $\tc$ and 
$\td$ lines are separated in region I by only horizontal
dimers and in region II by only vertical ones. 
Moreover, these separating dimers connect 
$\circ$ sites to $\bullet$ ones from left to right
in region I and from top to bottom in region II. 
We thus have 

\noindent{\bf Lemma 2.} {\sl All vertical dimers of the region I and
all horizontal dimers of region II are visited by $\tc$ or $\td$ 
lines. Also, all horizontal dimers
connecting $\bullet$ to $\circ$ sites from left to right in region I and
from top to bottom in region II are visited by $\tc$ or $\td$ lines.
}\par \noindent

% Fig 11
{\fig\inhell{
The immediate neighborhood  of L. Dimers are represented in red,
and fixed edges in light blue.  
Only two situations may occur for FPL paths:
either they cross L via a dimer on the left and a fixed horizontal edge on
the right, or they simply touch L without crossing via a fixed edge
followed by a dimer on the right side.
}{\epsfbox{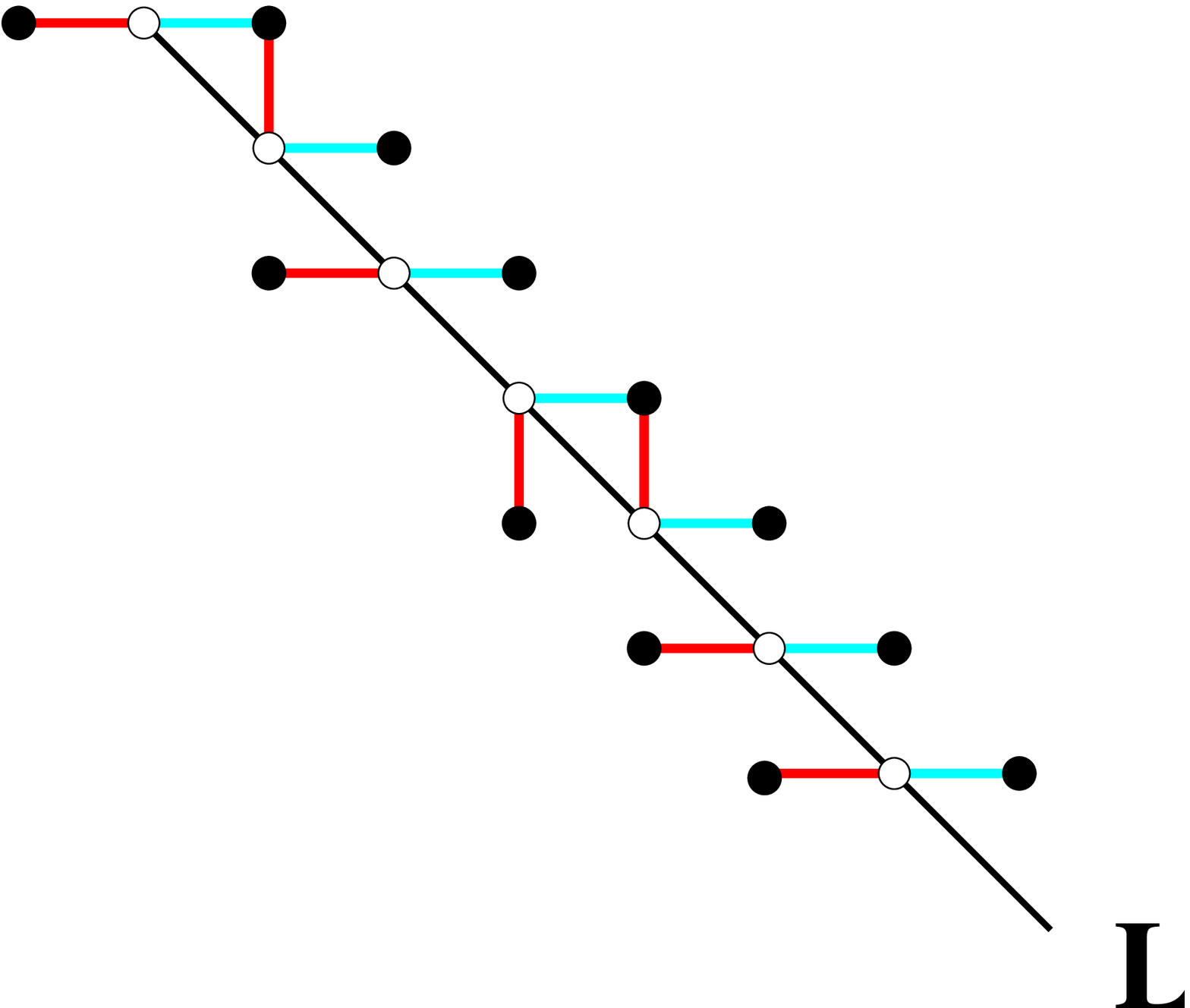}}}

\noindent{\bf Lemma 3.} {\sl All $\tc$ and $\td$ lines must go to L}.\nind
This is obvious on $\CD$: the $\tc$ and $\td$ lines enter $\CD$
through vertical edges along V; they must necessarily exit through
vertical edges again, and the only possibility to do so is through 
L$_\pm$, see Fig. \setting. From the standpoint of $\CO$, this is less
obvious: by parity arguments, all sites of $\CO$ 
on L are $\circ$ sites. By lemma 2, it follows that
any $\tc$ or $\td$ line reaching L must do so via a dimer. 
On Fig. \generic\ we observe that no fixed edge may touch L from the left
side hence the only way a FPL path may touch L from the left 
is via a dimer, which, as it ends up on a $\circ$ site, 
must by Lemma 2 be also part of
a $\tc$ or $\td$ line. So any FPL path that touches L from the left
and therefore crosses it, as it is prolongated via a fixed horizontal edge
on the right of L,
corresponds to the end of a $\tc$ or $\td$ line coming from the left.  
Analogously, any FPL path touching L from the right but not crossing it
corresponds to the end of a $\tc$ or $\td$ line coming from above. By
fully-packedness, only these two situations (crossing, or touching from the
right without crossing) may occur, in view of the disposition of fixed
edges, see Fig. \inhell\ for illustration.
On the total of $c+d$ sites of L, some of them, say $i$, correspond
to crossings
of FPL paths, and the other $c+d-i$ to touching without crossing.
These add up to the $c+d$ $\tc$ and $\td$ lines introduced above.

% Fig 12
{\fig\iequalc{Each  of the $i$ $\tc$ or $\td$ lines terminating on L from the left 
corresponds to an FPL arch centered on $C$, 
hence $i\leq c$. The same holds for
the $j$ $\tc$ or $\td$ lines touching L' from the left: each of them 
correspond to a FPL arch centered on $D$, 
hence $j\leq d$. But as $i+j=c+d$, we must have $i=c$ and $j=d$}{\epsfbox{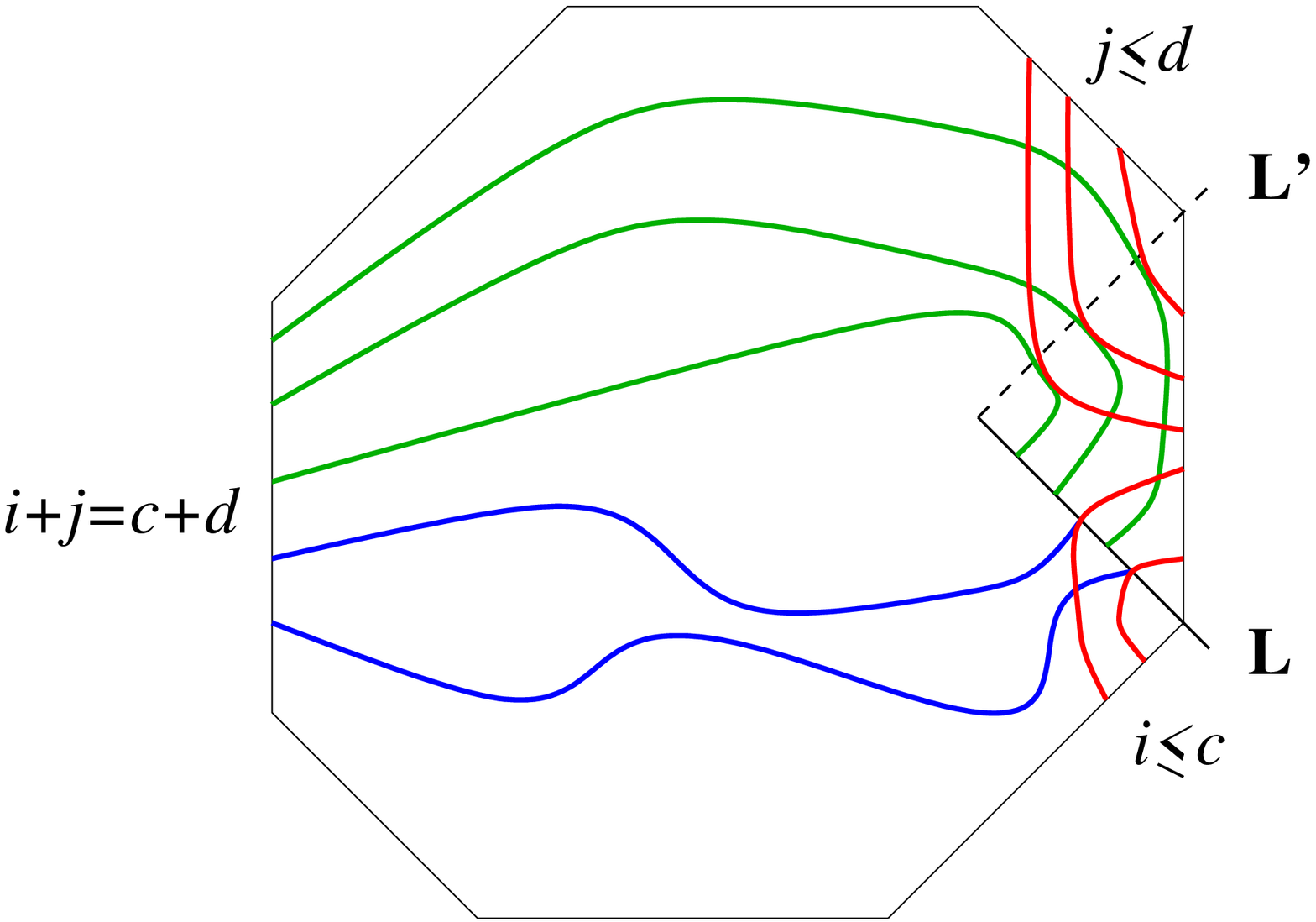}}}

\noindent{\bf Lemma 4.} {\sl $i=c$, hence all $\tc$ lines reach L from
the left and all $\td$ lines reach it from above.} 
\nind As noted above, the external endpoint of L is the center of the set 
of $c$ FPL nested arches. It is clear that these arches must cross
either L or its prolongation across the central square. Let us show that
they cross L and that they are in one-to one correspondence with the
$\tc$ or $\td$ lines ending up on L from the left. 
Let us prolongate the lines crossing L into FPL paths by letting them
alternate between fixed edges and dimers. The fixed edges on the right of L
being all horizontal, let us orient them from left to right. 
The FPL paths visit them in this direction. In particular, such a path can
never cross L again and can only bounce off it. This means that all
FPL paths crossing L must exit $\CO$  along its right border. The
same reasoning in the region to the left of L shows that these FPL paths
must enter the grid from the lower 
border. The $c$ FPL paths are the only ones
which connect the bottom to the right, 
and therefore $i\leq c$, see Fig.\iequalc. 
The same reasoning applied to the line
L', with $c$ replaced by $d$ shows that there are $j\leq d$ $\tc$ or
$\td$ lines
crossing L'. But the only way for a $\tc$ or $\td$ line to reach 
L from above is
by first crossing L', hence there are at most $j\leq d$ $\tc$ or $\td$ lines
reaching L from above. As the total of $\tc$ and $\td$ lines
reaching L is $c+d=i+j$, we must have $i=c$ and $j=d$.

\nind 
{\bf  Lemma 5. }{\sl From the  non-intersecting $\tc$ and $\td$ lines,
one reconstructs a unique  FPL
configuration of type $(a,b,c,d)$.}

\nind
We start from a configuration of non-intersecting $\tc$ and $\td$
lines going from the left 
vertical border V of $\CO$ to the line L, with the rule that every
second edge is horizontal and travelled from left to right
in region I and vertical and travelled from top to bottom in region II, while
the arrival sites on L form two disjoint sets of respectively $c$ and $d$
sites. We then construct dimers by keeping every 
second edge (those going from a $\bullet$ to a $\circ$)
on the $\tc$  and $\td$ lines, and adjoining them 
the horizontal edges connecting $\circ$ to $\bullet$ sites from left to
right in region I, and from top to bottom in region II,  in all regions
between the $\tc$ and/or $\td$ lines: this gives a complete
 dimer covering. 
Upon addition of the fixed edges as specified in sect. 1.1, 
one gets a FPL \conf. 
By the same argument as in Lemma 4, there are $c$ FPL 
paths crossing L and $d$ crossing L'. This exhausts all 
external edges on the right vertical side of the grid, 
and therefore the sets of $c$ and $d$ nested arches are next to one
another.
% Fig 13
{\fig\abcd{Artist's view of the
$\ta,\,\tb,\,\tc,\, \td$ lines on $\CO'$.}{\epsfbox{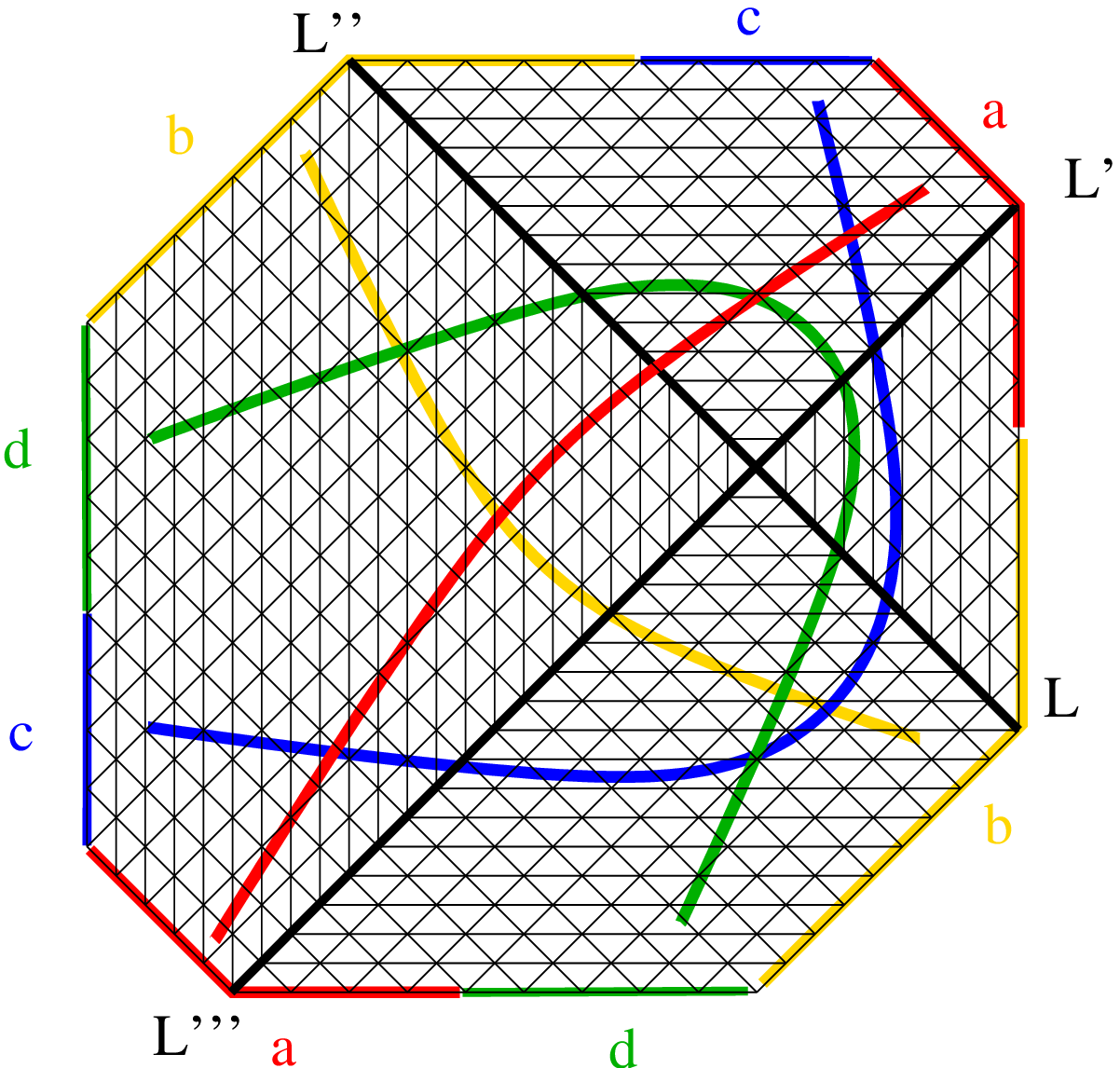}}}
The same discussion must now be repeated for the pairs $(d,a)$ and 
$(b,c)$. One introduces the two new sets of lines $\ta$ and $\tb$, 
which start from the segments of the boundary 
of $\CO'$ marked in Fig. \abcd, and which are de Bruijn lines
describing chains of tiles sharing edges parallel to a certain 
direction. One also continues the previous lines $\tc$ and $\td$
across the segment L up to the boundary. 
The segments L'' and {L'}'', which are the  continuations
of L and L', respectively, across the center of the central square, 
have lengths $a+d$ and $b+c$. 
One may then apply the 
analysis of Lemmas 1-5 to the pairs $(\ta,\td)$ and $(\tb,\tc)$. In 
particular, there are $a$ FPL paths  centered on $A$ which cross
the segment L'' on sites disjoint from those of the $\td$ lines; 
and likewise for the $\tb$ lines centered on $B$ and the $\tc$ lines, 
across the segment {L'}''.  One concludes that the FPL configuration 
contains at least four sets of $a$, $b$, $c$ and $d$ nested arches. 
As $a+b+c+d=n$, this exhausts the number of
FPL paths (and possibly also the patience of the reader!), 
establishes the Lemma and completes 
the proof the Theorem.

%%%%%%%%%%%%%%%%%%%%%%%%%%%%%%%%%%%%%%%%%%%%%%%%%%%%%%%%%%%%%%%%%%%%%%%%%%%%
%{\epsfxsize=1cm}
\def\fplmovev{${\epsfysize=5mm\epsfbox{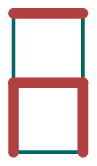}}\leftrightarrow
{\epsfysize=5mm\epsfbox{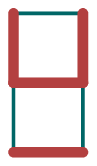}}$}
\def\fplmoveh{${\epsfxsize=5mm\epsfbox{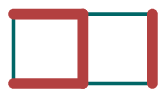}}\leftrightarrow
{\epsfxsize=5mm\epsfbox{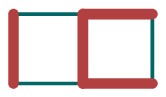}}$}
\def\ppmove{$\epsfxsize=5mm{\epsfbox{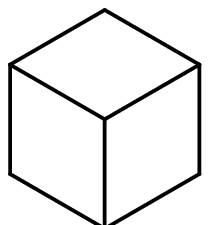}}\leftrightarrow
{\epsfxsize=5mm\epsfbox{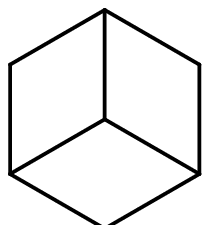}}$}

\def\nilmove{$\epsfxsize=5mm{\epsfbox{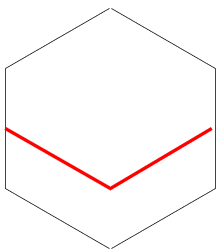}}\leftrightarrow
{\epsfxsize=5mm\epsfbox{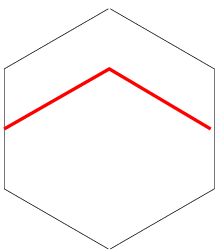}}$}
\def\sqmove{$\epsfxsize=3mm{\epsfbox{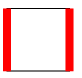}}\leftrightarrow
{\epsfxsize=3mm\epsfbox{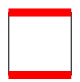}}$}

\subsec{ Remarks}
\nind 
1. From the discussion above, in particular from Lemma 5, it follows 
that

\noindent{\bf Corollary} {\sl Any dimer \conf\ on the octagonal grid
$\CO$ or any tiling of $\CO'$
yields a FPL of type $(a+p,b-p,c+p,d-p)$, for some $p$, 
$-c \le p \le d $. }

Indeed, start from the domino grid pertaining to any $(a,b,c,d)$ 
FPL \conf. For any dimer \conf\ on $\CO$ or the associated tiling
of $\CO'$,  draw the corresponding non intersecting lines on $\CD$. 
Let $c'$ be the number of those which reach L$_-$, $d'$ that of those 
reaching L$_+$, with $c+d=c'+d'$. By lemma 5, one reconstructs a 
unique FPL \conf\ of type $(a',b',c',d')$ with $a'-a=c'-c=b-b'=d'-d=:p$.

This corollary implies a sum rule on numbers 
of FPL configurations in terms of that of dimers
\eqn\sumrule{
\sum_{p=-\inf(a,c)}^{\inf(b,d)} \, A_n(a+p,b-p,c+p,d-p)=\#{\rm dimers\
on\ } \CO\ , }
of which we shall present examples in sect. 2.3.

\nind 2.
It is legitimate to ask if the elementary moves 
\fplmoveh\ or \fplmovev\ on dimers are ergodic, 
i.e. suffice to span all the FPL \conf s of a given type $(a,b,c,d)$.
Equivalently, are the moves \ppmove\ acting on tilings ergodic?
Contrary to the case of three sets of nested arches \PPJB, we cannot rely 
on the picture of cube stacking,
because of the conic singularity in the tiling caused by the cut.
The third picture we used, namely the non intersecting lines, provides 
the answer. It is easy to prove by  contradiction that all 
\conf s of non intersecting lines $\tc$ and $\td$ on the domain $\CD$
are generated from
one of them by repeated applications away from the branch point
of the elementary move \nilmove\ (and their rotated). 
This establishes the ergodicity property for the above moves of
dimers or of tiles
\foot{This result is stronger than the classical result stating that
 the moves \ppmove\ are ergodic on tilings of simply connected
domains of the triangular lattice 
%Indeed the latter applies only  to a domain of fixed shape, while 
%our result also involves alterations of $\CD$ along L${}_\pm$
\SalTo.}.
Note on the other hand that the move \sqmove\
acting on the central square connects FPL \conf s of types
$(a,b,c,d)$ and $(a\pm 1, b\mp 1, c\pm 1, d\mp 1)$.

\nind 3. In view of this ergodicity, 
we may now reconsider the two special configurations, ``empty'' and
``full'', depicted in Fig. \fullemptyd\ and \tilfulempt. 
What makes them extremal is the fact that only {\it two}
elementary moves can  act upon them. (This should be compared
with the unique move in the ordinary case of cube stacking, when one
considers the empty
or the full box.) The locations of these moves are what is represented by
the blue dots in Fig. \fullemptyd. 

\nind 
4. Note that the tiling problem we have to deal with is reminiscent of
but not identical to the problem treated in \DMB, namely that of a 
centro-symmetric octagon by six species of tiles. 
%%%%%%%%%%%%%%%%%%%%%%%%%%%%%%%%%%%%%%%%%%%%%%%%%%%%%%%%%%%%%%%%%%%%%%%%%%%

\newsec{Counting configurations.}

\nind In this section, we use 
the bijection of Theorem 1 to actually count the numbers of FPL
configurations of type $(a,b,c,d)$.

%\penalty -10000
\subsec{The setting}
%\penalty 10000
% Fig 14
{\fig\genercd{
A typical configuration of non-intersecting $\tc$ and $\td$ lines for $c=3$,
contributing to $f_{m_1,m_2,m_3}(a,b,3,d)$. The lines must enter through the
light blue points (on the left) and exit through the upper
part of the cut except at the points $m_1,m_2,m_3$ or through its lower part,
at the points $m_1,m_2,m_3$. 
The elementary steps   for the  $\tc$ and $\td$ lines,
in directions $\bf i$ and $\bf j$, are shown  in the medallion. 
}{\epsfbox{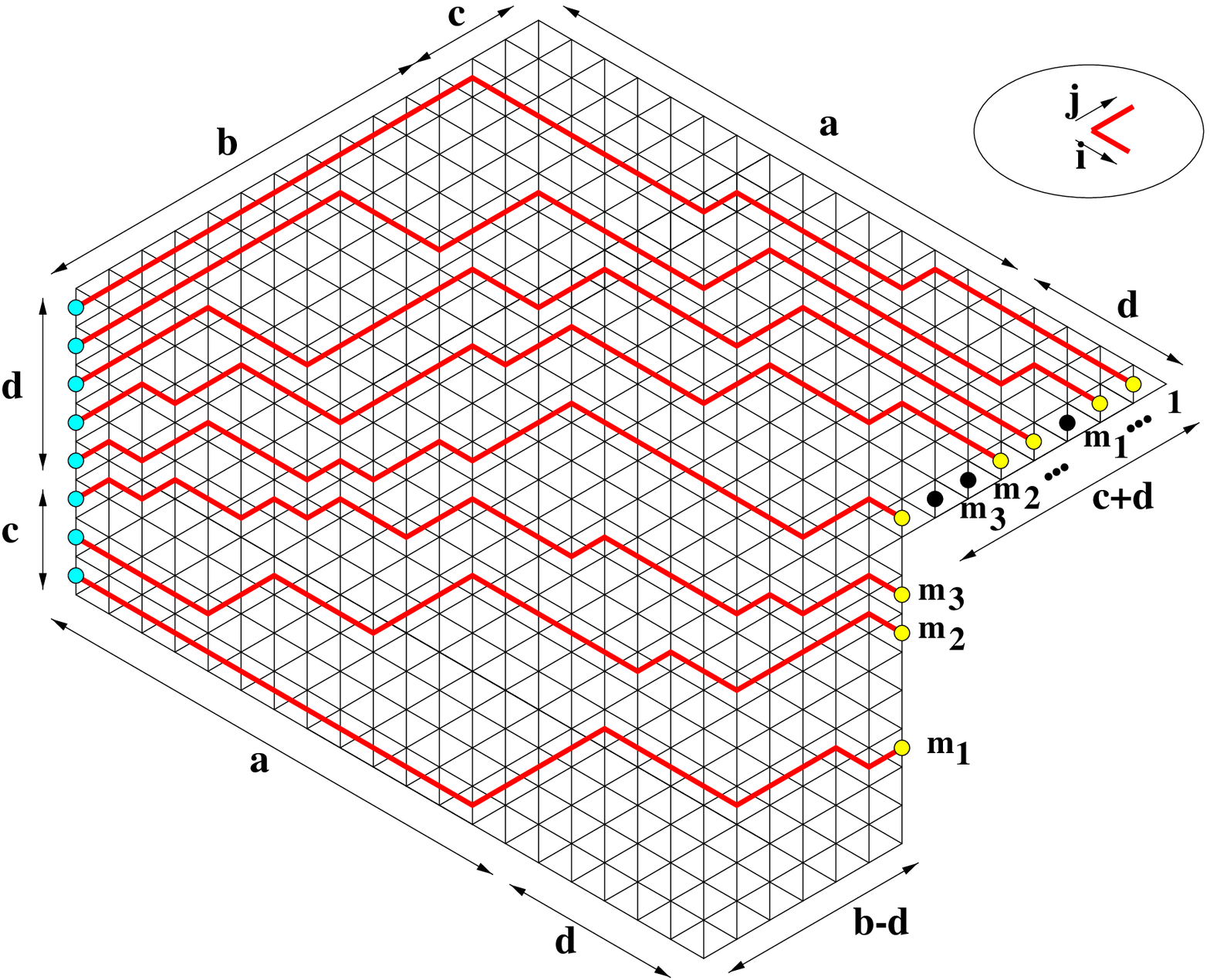}}}

It is simplest to use the $\tc$ and  $\td$ lines on the domain $\CD$.
The fact that they stop on segments L$_\pm$ reduces the counting of 
configurations
to the standard problem of enumerating all non-intersecting
paths from one set of points to another. As the tiles crossed by $\tc$
and $\td$ lines only
have two orientations  (and all have two vertical edges), 
the corresponding paths 
may only go in two directions say ${\bf i},{\bf j}$ at each point, as shown in 
the medallion of Fig.\genercd.

Moreover, we know that the $\tc$ lines 
end on the lower part of the cut L$_-$ 
at $c$ distinct points, marked in yellow in Fig.\genercd,
and that the $\td$ lines reach the upper part of the cut L$_+$
on $d$ distinct points, forming the complement of the $c$.
Let us denote by $m_1,m_2,...,m_c$ the positions of these $c$ points counted
from bottom to top on L$_-$, with $1\leq m_1<m_2<...<m_c\leq c+d$,
while their positions must be counted from top to bottom on L$_+$.

We now have the well-posed problem of computing the number of
configurations of non-intersecting $\tc$  and $\td$ lines with elementary
steps $\bf i$ or $\bf j$ going on $\CD$ from its left side  (light blue 
points in Fig.\genercd) to its right 
(yellow points in Fig.\genercd), and such that the lower $c$ lines
exit through points $m_1,m_2,...,m_c$ of the lower part of the cut, 
while the upper d ones
exit through their $d$ complements on the upper part of the cut.
We denote by $f_{m_1,m_2,...,m_c}(a,b,c,d)$ the number of such configurations.
 
\subsec{A fermionic formula for path counting}

\nind 
In this section we recall the so-called Lindstr\"om-Gessel-Viennot \LGV\
formula for counting 
the number $N(A_1,A_2,...,A_N\vert E_1,E_2,...,E_N)$ of non-intersecting paths
from a set of points $A_1,A_2,...,A_N$ to a set of points 
$E_1,E_2,...,E_N$ of the square
lattice, such that only elementary steps to the right and to the top are allowed.
Let $P(A_i,E_j)$ denote the number of paths from the point $A_i$ to the point $E_j$
with these only allowed elementary steps, then we have (see \LGV\ and
further references in \CK)
\eqn\lgv{ N(A_1,A_2,...,A_N\vert E_1,E_2,...,E_N)
=\det\left(P(A_i,E_j)\right)_{1\leq i, j\leq N}}
(For physicists, recall this is just the expression of the $2N$-point 
function $\langle \psi_{A_1}...\psi_{A_N}\psi_{E_1}^*
... \psi_{E_N}^*\rangle$ of free fermions, expressed 
as the Slater determinant of their 2-point function $P(A_i,E_j)=\langle \psi_{A_i}
\psi^*_{E_j}\rangle$, when their action is
$\sum_{\alpha\to\beta}\psi^*_\alpha \psi_\beta$,  
summed over all edges $\alpha\to \beta$ of the square
lattice,  oriented from left to right and from bottom to top.)

%%%%%%%%%%%%%%%%%%%%%%%%%%%%%%%%%%%%%%%%%%%%%%%%%%%%%%%%%%%%%%%%%%%%%%%

\subsec{ Computation of $A_n(a,b,c,d)$}

\nind
To apply formula \lgv\ to the computation of
$f_{m_1,..,m_c}(a,b,c,d)$,  
we  just have to record the relative
positions of the entry and exit points of the $\tc$ and $\td$ lines. 
The solution goes as follows.
We first construct the matrix $\cal M$ of size $(c+d)\times (c+d)$ 
whose ${\cal M}_{i,j}$ entry is the binomial coefficient
\eqn\entrim{ {\cal M}_{i,j}={a+b+c+d-j\choose a+d-i}\ . }
The matrix element ${\cal M}_{i,j}$ counts generically the number of
paths  between the $i$-th 
entry point counted from bottom to top (in light blue 
in Fig.\genercd) and 
the $j$-th exit, counted from top to bottom on the upper part of the
cut  (in yellow in Fig.\genercd).
To take into account the missing 
images of the points on the lower
part of the cut (filled black dots in Fig.\genercd), we must 
erase the columns $j=m_1,m_2,...,m_c$ of this matrix, and append 
instead $c$ new columns $v_k$, 
%replace the columns $j=m_1,m_2,...,m_c$ of the matrix  ${\cal M}$ by the $c$
%columns $v_k$, 
$k=1,2,...,c$, %all of size $(c+d)\times 1$, and with entries
\eqn\addcol{ (v_k)_i= {a+b\choose a+c+2d+1-i-m_k}\,, }
which count the total number of paths from the $i$-th entry point to 
the exit point $m_k$ on the lower part of the cut. Then  the determinant
of the resulting matrix ${\cal M}(m_1,m_2,...,m_c)$ gives $f_{m_1,..,m_c}(a,b,c,d)$:
\eqn\lutfin{ 
f_{m_1,..,m_c}(a,b,c,d)=\det\left({\cal M}(m_1,m_2,...,m_c)\right) }
and finally 
\eqn\luttefinale{ A_n(a,b,c,d)=\sum_{1\leq m_1<m_2<\cdots <m_c\leq c+d}
f_{m_1,m_2,\cdots,m_c}(a,b,c,d)}

This formula is very explicit if not too easy to manipulate. 
We have been able to drastically
simplify it in a few cases. For instance, when $c=1$, 
one may expand  $\det{\cal M}(m)$  with respect to its added
column, and use an explicit expression for
 the inverse matrix of ${\cal M}$ to prove that
\eqn\generfm{\eqalign{
f_m(a,b,1,d)&={{b\choose d+1-m}{d\choose m-1}\over
{a+2d+1-m\choose a}{2d+1-m\choose m-1}{b+d+1\choose m-1}}\
{\prod_{i=0}^d {a+b+i\choose a}\over
\prod_{i=0}^{d-1} {a+i\choose a}}
\times\cr
&\ \times\ \sum_{j=0}^{m-1}
{{2d+2-m\choose m-1-j}
{2d+2+j-2m\choose j}{a+m-2-j\choose m-1-j}{b+d+2+j-m\choose j}
\over {m-1\choose j}}
\cr}\ .}
For $c=2$ we have
\eqn\twocs{\eqalign{
&f_{m_1,m_2}(a,b,2,d)={{b\choose d+1-m_1}{b+1\choose d+2-m_2}
{d+1\choose m_1-1}{d\choose m_2-2}\over
{a+2d+2-m_1\choose a}{a+2d+2-m_2\choose a}
{b+d+3\choose m_1-1}{b+d+2\choose m_2-2}}\
{\prod_{i=0}^{d+1} {a+b+i\choose a}\over
\prod_{i=0}^{d-1} {a+i\choose a}}
\times\cr
&\ \times\ \sum_{j_1=0}^{m_1-1}\sum_{j_2=0}^{Min(m_2-2,m_2-m_1+j_1)}
\left( 1-{{j_2\choose m_2-m_1}\over {j_1+m_2-m_1\choose m_2-m_1}}\right)
{(m_2 - m_1) (m_2 - m_1 - (j_2 - j_1))\over (m_2 - 1 - j_2)(d+2-m_1)}\times \cr
&\times {(2d+4-m_1)(2d+3-m_1)(2d+4-m_2)(2d+3-m_2)\over (2d+5+j_1-2m_1)
(2d+5+j_2-2m_2)(2d+5+j_1-m_1-m_2)(2d+5+j_2-m_1-m_2)}\times \cr
&\times {a-3+m_1-j_1\choose a-2}{a-3+m_2-j_2\choose a-1}{b+d+4+j_1-m_1\choose j_1}
{b+d+4+j_2-m_2\choose j_2}\cr}\ .} 
This follows from a double expansion of $\det{\cal M}(m_1,m_2)$
with respect to the two added columns.

Both expressions  yield
 pretty explicit formulas for $N(a,b,1,d)$ and $N(a,b,2,d)$.

%
%%%%%%%%%%%%%%%%%%%%%%%%%%%%%%%%%%%%%%%%%%%%%%%%%%%%%%%%%%%%%%%%%%%%%%
%\bye

\subsec{FPL and dimer countings}

\nind According to the third remark of sect. 1.4 and eq. \sumrule, 
it may be interesting to compute also the total number of dimers
on the octagonal grid $\CO$. 

% Fig xxx  % {\def\epsfsize#1#2{0.6#1}\fig\xxx{bla bla}{\epsfbox{xxx.eps}}}
% Fig 15
{\fig\dimer{Dimer counting : the case
(1,1,1,1). }{\epsfbox{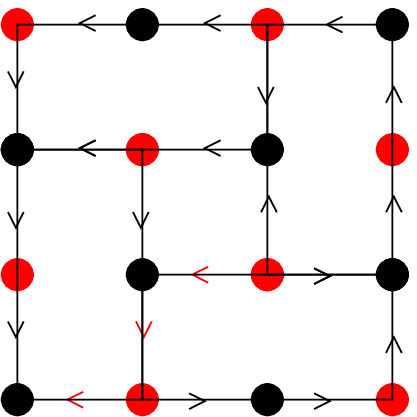}}}
We shall illustrate this by an example. 
In the cases of an $(m,1,m,1)$ FPL, the domino grid $\CO$ is
made of a unit square surrounded by rectangular dominos. 
According to Kasteleyn \Kas, one computes the number of dimers
on this graph by orienting its edges in such a way 
that along every elementary closed circuit the number of edges of either
orientation is odd.  The number of dimers is then the 
pfaffian of the resulting signed adjacency matrix. 
The case $m=1$  is depicted in Fig. \dimer.

Since these graphs are 2-colored,  their
(antisymmetric) signed adjacency matrix is made of two off-diagonal blocks, 
$$ \pmatrix{0 & G_m\cr -G_m^T & 0\cr}$$ and the pfaffian 
is just  (up to a sign) the determinant of one of these
blocks.  (It is 9 for the case  $(1,1,1,1)$).
{\fig\constr{Constructing $G_{m+1}$ from $G_m$ 
(in black) by adding one layer of (blue) dominos.}{\epsfbox{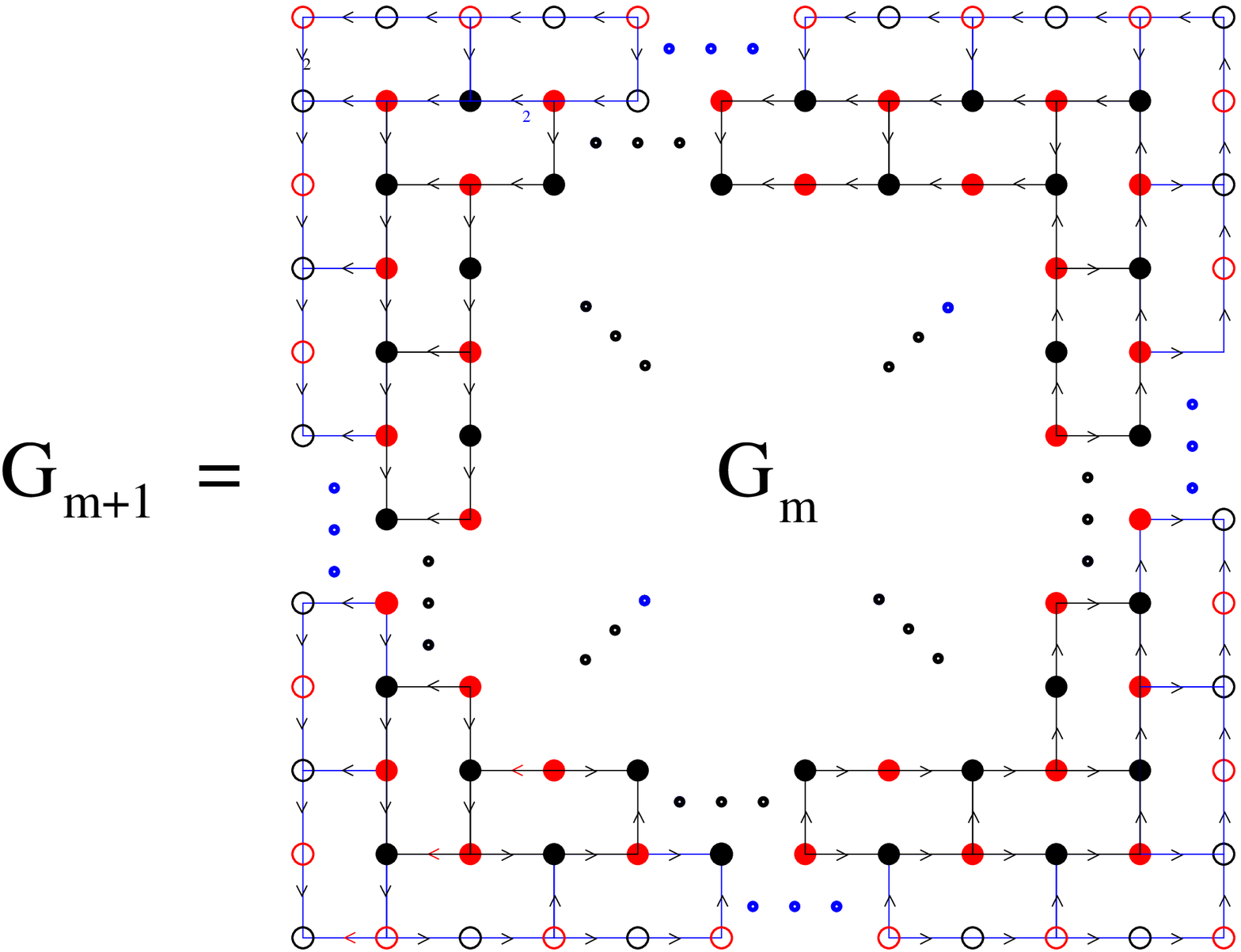}}}
In general, one constructs  the graphs and their adjacency matrix 
by  a recursive procedure of adding layers of dominos, as depicted
in Fig. \constr.

One finds that the determinants $D_m$, thus the numbers of 
dimers, are alternatingly perfect squares or doubles of perfect
squares. 
This is a consequence of the 4-fold symmetry of the grid, as discussed 
in \Jock.
(In the terminology of that paper, the graphs $G_m$ are 
4-odd-symmetric, and their $\Bbb{Z}_4$ quotient has $(m+1)^2$ vertices.)

$$\eqalign{& d_m:=\sqrt{D_{2m-1}}=
3,70,13167,20048886,247358122583,24736951705389664, \cr &
  20054892679528741176540,131821539275853806053297420440, \cdots
\omit{\cr
&  7025331201927872335198652663701963740, \cr
& 3035820544690570843477213690226325544429730520,\cr
&10637121157772642319653355663448614818452672194577130535, \cr
30221626628&2282935095067107120753635189646893208871258726798741600, \cr
6962455612123691314621&7318499574924313015409403887855745929020536287805987932,\cr
130065411959896416903405918670&348834958202869510082283238601175000873412814426
099181498624\cr }
\qquad {\rm for\ } m=1,2,\cdots, 8 %14
\cr}$$
while
$$\eqalign{& \delta_m:=\sqrt{D_{2m}/2}=
8,
  526,
  280772,
  1215446794,
  42663813089328, \cr
&  12142696908022734304,
  28022410984084414473869168,\cr
 & 524367885668519092847372976461256, \cdots 
\omit{\cr
&  79563171848625690733121191356810869527032, \cr
& 97889908109610487125047187380445257609818946799364, \cr
976601&143695886505725028838695361888364280057606376867851408, \cr
7900471236761217965&6475785565442541556601008586938332140945884608691616, \cr
5182599787046516921248731909235&9184143291360389647774243217930195054571611409544672\cr }
\qquad {\rm also\ for\ } m=1,2,\cdots, 8 %14. 
}$$
%{3, 70, 13167, 20048886, 247358122583, 24736951705389664, \
%20054892679528741176540, 131821539275853806053297420440, \
%7025331201927872335198652663701963740, \
%3035820544690570843477213690226325544429730520, \
%10637121157772642319653355663448614818452672194577130535, \
%302216266282282935095067107120753635189646893208871258726798741600, \
%69624556121236913146217318499574924313015409403887855745929020536287805987932,\
% 130065411959896416903405918670348834958202869510082283238601175000873412814426099181498624}

%(Many more terms have been computed but are omitted here for brevity.)
 
%
Curiously, one observes \foot{We are very grateful to Saibal Mitra
for this observation.} that these numbers are given by the determinant
$\hat D(L,\theta):= e^{-i\theta L/2}
\det_{1\le i.,j\le L} \left( {i+j-2\choose i-1}+e^{i\theta} \delta_{ij}\right)$
for $\theta=\pi/2$, namely that $\sqrt{D_{m}}=\hat D(m+1,\pi/2)$.
This may be proved by writing ${\hat D(m+1,\pi/2)}{\hat D(m+1,-\pi/2)}=\det(I+T^2)$,
where $T$ is the matrix with entries $T_{ij}={i+j-2\choose i-1}$, $i,j=1,2,...,m+1$,
and expanding the determinant in terms of multiple  minors of $T^2$. More precisely,
introducing $\Delta_m(\lambda)=\det(\lambda I+T^2)$, we have the expansion
\eqn\introdm{\Delta_m(\lambda)=\sum_{k=0}^{m+1} \lambda^{m+1-k} 
\sum_{1\leq p_1<...<p_k\leq m+1} \det_{1\leq i,j\leq k}\left( (T^2)_{p_i,p_j} \right)}
The coefficient of $\lambda^{m+1-k}$ in this polynomial is then identified with the sum
of determinants
\luttefinale, with $a=m+1-k,b=k,c=m+1-k,d=k$, by expanding the latter with respect
to some of their columns as follows. Upon the redefinition of columns $j\to m+2-j$, we note 
that the matrix ${\cal M}_{ij}\to {\cal M}'_{ij}={m+j\choose m+1-i}$, while the added columns
are borrowed from the matrix ${\cal V}'_{ij}={m+1\choose m+1+j-i}$,
which is lower triangular with $1$'s on the diagonal. 
The matrix 
${\cal M}(m_1'=m+2-m_1,...,m_{m+1-k}'=m+2-m_{m+1-k})$ is now made of the matrix 
${\cal M}'$ with columns $m_1',m_2',...,m_{m+1-k}'$ 
erased and those of ${\cal V}'$ appended.
%replaced by those of  ${\cal V}'$. 
Denoting by $p_1,p_2,...,p_k$ the ordered complement of the $m'_i$'s in $\{1,2,...,m+1\}$,
we may also view this matrix as made of ${\cal V}'$, with the columns 
$p_1,...,p_k$
erased and those of ${\cal M}'$ appended.
%replaced by those of  ${\cal M}'$. 
Expanding the corresponding determinant with respect to 
these columns, we arrive at
\eqn\arrive{ \det({\cal M}(m_1',...,m_{m+1-k}'))=
\sum_{1\leq i_1,...,i_k\leq m+1} 
(-1)^{\sum m+1+i_r}{\cal M}'_{i_1,p_1}...{\cal M}'_{i_k,p_k} 
|{\cal V}'|_{i_1,...,i_k;p_1,...,p_k} }
where the latter denotes the multiple minor of ${\cal V}'$ in which 
lines $i_1,...,i_k$
and columns $p_1,...,p_k$ are erased and the determinant is taken.
The latter is  also equal to
the determinant of single minors $\det(|{\cal V}'|_{i_r,p_s})_{1\leq r,s\leq k}$,
by a property satisfied by all lower triangular matrices with $1$'s on the diagonal.
Using the multilinearity of the determinant, we write
\eqn\wewrite{  \det({\cal M}(m_1',...,m_{m+1-k}'))=\det_{1\leq r,s\leq k}\left(
\sum_{i=1}^{m+1} (-1)^{m+1+i} {\cal M}'_{i,p_r}|{\cal V}'|_{i,p_s}\right) }
and we finally note that 
\eqn\notethat{ \sum_{i=1}^{m+1} (-1)^{m+1+i} {\cal M}'_{i,r}|{\cal V}'|_{i,s}=T^2_{rs}}
as a consequence of a simple binomial identity, as $|{\cal V}'|_{i,s}={m+s-i-1\choose m}$.
This completes the proof, as the
sum over the $m_i'$ amounts to that over the $p_i$. 

So we get all the $A_{2m+2}(a,b,c,d)$ with 
$a+b=b+c=c+d=m+1$ as coefficients of the polynomial $\Delta_m(\lambda)$:
\eqn\finlut{\Delta_m(\lambda)=\det(\lambda I+T^2)=\sum_{k=0}^{m+1} \lambda^{m+1-k} 
A_{2m+2}(k,m+1-k,k,m+1-k) }
with the convention that $A_n(a,0,a,0)=1=A_n(0,a,0,a)$, as it counts the number
of FPL configurations with a single set of nested arches.
Moreover, according to Corollary 1, the numbers $D_m$ are nothing but
\eqn\check{ D_m= \sum_{k=0}^{m+1} A_{2m+2}(k,m+1-k,k,m+1-k)=\Delta_m(1)}
One may illustrate
\finlut-\check\ on the first values of $m$
$$\eqalignno{ m&=1:\qquad T^2=\pmatrix{2 & 3\cr 3 & 5}
\qquad \Delta_1(\lambda)=1+7\lambda+\lambda^2\cr 
D_1&= 9= 1+ A_4(1,1,1,1)+1= 1+7+1 \cr
m&=2:\qquad T^2=\pmatrix{3 & 6 & 10\cr 6 & 14 & 25\cr 10 & 25 & 46}
\qquad \Delta_2(\lambda)=1+63\lambda+63\lambda^2+\lambda^3\cr
D_2&= 128= 1+ A_6(2,1,2,1)+A_6(1,2,1,2)+1= 1+63+63+1 \cr
m&=3:\qquad T^2=\pmatrix{4 & 10 & 20 & 35\cr 10 & 30 & 65 & 119\cr 20 & 65 & 146 & 273\cr
35 & 119 & 273 & 517}\qquad \Delta_3(\lambda)
=1+697\lambda+3504\lambda^2 +697\lambda^3+\lambda^4\cr
D_3&=4900=1+A_8(3,1,3,1)+A_8(2,2,2,2)+A_8(1,3,1,3)+1=1+697+3504 +697+1 \cr
{\rm etc.} }$$

Note that the above proof relies crucially 
on the fact that ${\cal V}'$ is lower triangular with $1$'s on the diagonal, 
a property still true in general
when $b=d$, while $a$ and $c$ are arbitrary (in which case we have
${\cal V}'_{ij}={a+b\choose a+b+j-i}$). 
This leads straightforwardly to the generating
function:
\eqn\straitfo{
\Delta_{a+b,b+c}(\lambda)= 
\det_{1\leq i,j\leq b+c}\big(\lambda I+ U(a+b)\big)=
\sum_{i=0}^{{\rm min}(a+b,b+c)}
\lambda^{b+c-i} A_n(a+b-i,i,b+c-i,i) }
where the matrix $U(m)$ is the $m$-truncated version of $T^2$, namely
\eqn\ntrunc{ U(m)_{ij}=\sum_{r=1}^m {i+r-2\choose i-1}{j+r-2\choose j-1} }
This gives access to all FPL numbers of the form $A_n(a,b,c,b)$ in a very compact manner.
For illustration, we have for $a+b=4$ and $b+c=6$ the following generating function
\eqn\geneillu{\eqalign{ \Delta_{4,6}(\lambda)&=\det\pmatrix{
\lambda+4 & 10 & 20 & 35 & 56 & 84\cr
10 & \lambda+30 & 65 & 119 & 196 & 300\cr
20 & 65 & \lambda+146 & 273 & 456 & 705 \cr
35 & 119 & 273 & \lambda+517 & 871 & 1355\cr
56 & 196 & 456 & 871 & \lambda+1476 & 2306\cr
84 & 300 & 705 & 1355 & 2306 & \lambda+3614}\cr
&=\lambda^6+5787\lambda^5+129627\lambda^4+97874\lambda^3+1764\lambda^2\cr
&=\lambda^6+A_{10}(3,1,5,1)\lambda^5+A_{10}(2,2,4,2)\lambda^4
+A_{10}(1,3,3,3)\lambda^3+A_{10}(0,4,2,4)\lambda^2\cr}}
and the corresponding number of dimer configurations is 
$\Delta_{4,6}(1)=235053$.

The determinants $\hat D(L,\theta)$ have occurred in different contexts 
\refs{\CEKZ,\MNosc,\MNnew}. In the latter of these references, 
the following large $L$ asymptotic behavior is proposed:
$$ \hat D(L,\theta)\approx A_{{\rm HT}}(L)^2\approx \left({3^3\over 2^4}\right)^{
L^2/4}\ .$$
%

%%%%%%%%%%%%%%%%%%%%%%%%%%%%%%%%%%%%%%%%%%%%%%%%%%%%%%%%%%%%%%%%%%%%%%%%%%%
\bigskip

\nind{\bf Acknowledgements. } \nind 
Many thanks to
Christian Krattenthaler for his insight and his many suggestions
of simplifications of the arguments of sect. 1 and
of the computation of sect. 2, 
Saibal Mitra for a crucial observation, 
R\'emy Mosseri for discussions, 
and Paul Zinn-Justin for his wonderful software, accessible on %\nind
{\tt http://ipnweb.in2p3.fr/$\sim$lptms/membres/pzinn/fpl} \ .\nind 
This work is partially supported by the European networks
HPRN-CT-2002-00325 and HPRN-CT-1999-00161.

%%%%%%%%%%%%%%%%%%%%%%%%%%%%%%%%%%%%%%%%%%%%%%%%%%%%%%%%%%%%%%%%%%%%%%%%%%%

\listrefs

\end

%% file: mssymb.tex
%               *****     MSSYMB.TeX    *****                  8 Jul 87

%

%       This file contains the definitions for the symbols in the two

%       "extra symbols" fonts created at the American Math. Society.

%

%       Ce fichier a 't' modifi' le 20 Septembre 1991, pour remplacer

%       les anciennes fontes msxm et msym par les fontes plus r'centes

%       msam et msbm. Le nom du fichier n'a pas 't' chang'.

\catcode`\@=11

\font\tenmsa=msam10

\font\sevenmsa=msam7

\font\fivemsa=msam5

\font\tenmsb=msbm10

\font\sevenmsb=msbm7

\font\fivemsb=msbm5

\newfam\msafam

\newfam\msbfam

\textfont\msafam=\tenmsa  \scriptfont\msafam=\sevenmsa

  \scriptscriptfont\msafam=\fivemsa

\textfont\msbfam=\tenmsb  \scriptfont\msbfam=\sevenmsb

  \scriptscriptfont\msbfam=\fivemsb

\def\hexnumber@#1{\ifcase#1 0\or1\or2\or3\or4\or5\or6\or7\or8\or9\or

	A\or B\or C\or D\or E\or F\fi }

%  The following 13 lines establish the use of the Euler Fraktur font.

%  To use this font, remove % from beginning of these lines.

\font\teneuf=eufm10

\font\seveneuf=eufm7

\font\fiveeuf=eufm5

\newfam\euffam

\textfont\euffam=\teneuf

\scriptfont\euffam=\seveneuf

\scriptscriptfont\euffam=\fiveeuf

\def\frak{\ifmmode\let\next\frak@\else

 \def\next{\Err@{Use \string\frak\space only in math mode}}\fi\next}

\def\goth{\ifmmode\let\next\frak@\else

 \def\next{\Err@{Use \string\goth\space only in math mode}}\fi\next}

\def\frak@#1{{\frak@@{#1}}}

\def\frak@@#1{\fam\euffam#1}

%  End definition of Euler Fraktur font.

\edef\msa@{\hexnumber@\msafam}

\edef\msb@{\hexnumber@\msbfam}

\mathchardef\boxdot="2\msa@00

\mathchardef\boxplus="2\msa@01

\mathchardef\boxtimes="2\msa@02

\mathchardef\square="0\msa@03

\mathchardef\blacksquare="0\msa@04

\mathchardef\centerdot="2\msa@05

\mathchardef\lozenge="0\msa@06

\mathchardef\blacklozenge="0\msa@07

\mathchardef\circlearrowright="3\msa@08

\mathchardef\circlearrowleft="3\msa@09

\mathchardef\rightleftharpoons="3\msa@0A

\mathchardef\leftrightharpoons="3\msa@0B

\mathchardef\boxminus="2\msa@0C

\mathchardef\Vdash="3\msa@0D

\mathchardef\Vvdash="3\msa@0E

\mathchardef\vDash="3\msa@0F

\mathchardef\twoheadrightarrow="3\msa@10

\mathchardef\twoheadleftarrow="3\msa@11

\mathchardef\leftleftarrows="3\msa@12

\mathchardef\rightrightarrows="3\msa@13

\mathchardef\upuparrows="3\msa@14

\mathchardef\downdownarrows="3\msa@15

\mathchardef\upharpoonright="3\msa@16

\mathchardef\downharpoonright="3\msa@17

\mathchardef\upharpoonleft="3\msa@18

\mathchardef\downharpoonleft="3\msa@19

\mathchardef\rightarrowtail="3\msa@1A

\mathchardef\leftarrowtail="3\msa@1B

\mathchardef\leftrightarrows="3\msa@1C

\mathchardef\rightleftarrows="3\msa@1D

\mathchardef\Lsh="3\msa@1E

\mathchardef\Rsh="3\msa@1F

\mathchardef\rightsquigarrow="3\msa@20

\mathchardef\leftrightsquigarrow="3\msa@21

\mathchardef\looparrowleft="3\msa@22

\mathchardef\looparrowright="3\msa@23

\mathchardef\circeq="3\msa@24

\mathchardef\succsim="3\msa@25

\mathchardef\gtrsim="3\msa@26

\mathchardef\gtrapprox="3\msa@27

\mathchardef\multimap="3\msa@28

\mathchardef\therefore="3\msa@29

\mathchardef\because="3\msa@2A

\mathchardef\doteqdot="3\msa@2B

\mathchardef\triangleq="3\msa@2C

\mathchardef\precsim="3\msa@2D

\mathchardef\lesssim="3\msa@2E

\mathchardef\lessapprox="3\msa@2F

\mathchardef\eqslantless="3\msa@30

\mathchardef\eqslantgtr="3\msa@31

\mathchardef\curlyeqprec="3\msa@32

\mathchardef\curlyeqsucc="3\msa@33

\mathchardef\preccurlyeq="3\msa@34

\mathchardef\leqq="3\msa@35

\mathchardef\leqslant="3\msa@36

\mathchardef\lessgtr="3\msa@37

\mathchardef\backprime="0\msa@38

\mathchardef\risingdotseq="3\msa@3A

\mathchardef\fallingdotseq="3\msa@3B

\mathchardef\succcurlyeq="3\msa@3C

\mathchardef\geqq="3\msa@3D

\mathchardef\geqslant="3\msa@3E

\mathchardef\gtrless="3\msa@3F

\mathchardef\sqsubset="3\msa@40

\mathchardef\sqsupset="3\msa@41

\mathchardef\vartriangleright="3\msa@42

\mathchardef\vartriangleleft="3\msa@43

\mathchardef\trianglerighteq="3\msa@44

\mathchardef\trianglelefteq="3\msa@45

\mathchardef\bigstar="0\msa@46

\mathchardef\between="3\msa@47

\mathchardef\blacktriangledown="0\msa@48

\mathchardef\blacktriangleright="3\msa@49

\mathchardef\blacktriangleleft="3\msa@4A

\mathchardef\vartriangle="0\msa@4D

\mathchardef\blacktriangle="0\msa@4E

\mathchardef\triangledown="0\msa@4F

\mathchardef\eqcirc="3\msa@50

\mathchardef\lesseqgtr="3\msa@51

\mathchardef\gtreqless="3\msa@52

\mathchardef\lesseqqgtr="3\msa@53

\mathchardef\gtreqqless="3\msa@54

\mathchardef\Rrightarrow="3\msa@56

\mathchardef\Lleftarrow="3\msa@57

\mathchardef\veebar="2\msa@59

\mathchardef\barwedge="2\msa@5A

\mathchardef\doublebarwedge="2\msa@5B

\mathchardef\angle="0\msa@5C

\mathchardef\measuredangle="0\msa@5D

\mathchardef\sphericalangle="0\msa@5E

\mathchardef\varpropto="3\msa@5F

\mathchardef\smallsmile="3\msa@60

\mathchardef\smallfrown="3\msa@61

\mathchardef\Subset="3\msa@62

\mathchardef\Supset="3\msa@63

\mathchardef\Cup="2\msa@64

\mathchardef\Cap="2\msa@65

\mathchardef\curlywedge="2\msa@66

\mathchardef\curlyvee="2\msa@67

\mathchardef\leftthreetimes="2\msa@68

\mathchardef\rightthreetimes="2\msa@69

\mathchardef\subseteqq="3\msa@6A

\mathchardef\supseteqq="3\msa@6B

\mathchardef\bumpeq="3\msa@6C

\mathchardef\Bumpeq="3\msa@6D

\mathchardef\lll="3\msa@6E

\mathchardef\ggg="3\msa@6F

\mathchardef\circledS="0\msa@73

\mathchardef\pitchfork="3\msa@74

\mathchardef\dotplus="2\msa@75

\mathchardef\backsim="3\msa@76

\mathchardef\backsimeq="3\msa@77

\mathchardef\complement="0\msa@7B

\mathchardef\intercal="2\msa@7C

\mathchardef\circledcirc="2\msa@7D

\mathchardef\circledast="2\msa@7E

\mathchardef\circleddash="2\msa@7F

\def\ulcorner{\delimiter"4\msa@70\msa@70 }

\def\urcorner{\delimiter"5\msa@71\msa@71 }

\def\llcorner{\delimiter"4\msa@78\msa@78 }

\def\lrcorner{\delimiter"5\msa@79\msa@79 }

\def\yen{\mathhexbox\msa@55 }

\def\checkmark{\mathhexbox\msa@58 }

\def\circledR{\mathhexbox\msa@72 }

\def\maltese{\mathhexbox\msa@7A }

\mathchardef\lvertneqq="3\msb@00

\mathchardef\gvertneqq="3\msb@01

\mathchardef\nleq="3\msb@02

\mathchardef\ngeq="3\msb@03

\mathchardef\nless="3\msb@04

\mathchardef\ngtr="3\msb@05

\mathchardef\nprec="3\msb@06

\mathchardef\nsucc="3\msb@07

\mathchardef\lneqq="3\msb@08

\mathchardef\gneqq="3\msb@09

\mathchardef\nleqslant="3\msb@0A

\mathchardef\ngeqslant="3\msb@0B

\mathchardef\lneq="3\msb@0C

\mathchardef\gneq="3\msb@0D

\mathchardef\npreceq="3\msb@0E

\mathchardef\nsucceq="3\msb@0F

\mathchardef\precnsim="3\msb@10

\mathchardef\succnsim="3\msb@11

\mathchardef\lnsim="3\msb@12

\mathchardef\gnsim="3\msb@13

\mathchardef\nleqq="3\msb@14

\mathchardef\ngeqq="3\msb@15

\mathchardef\precneqq="3\msb@16

\mathchardef\succneqq="3\msb@17

\mathchardef\precnapprox="3\msb@18

\mathchardef\succnapprox="3\msb@19

\mathchardef\lnapprox="3\msb@1A

\mathchardef\gnapprox="3\msb@1B

\mathchardef\nsim="3\msb@1C

%\mathchardef\napprox="3\msb@1D

\mathchardef\ncong="3\msb@1D

\mathchardef\varsubsetneq="3\msb@20

\mathchardef\varsupsetneq="3\msb@21

\mathchardef\nsubseteqq="3\msb@22

\mathchardef\nsupseteqq="3\msb@23

\mathchardef\subsetneqq="3\msb@24

\mathchardef\supsetneqq="3\msb@25

\mathchardef\varsubsetneqq="3\msb@26

\mathchardef\varsupsetneqq="3\msb@27

\mathchardef\subsetneq="3\msb@28

\mathchardef\supsetneq="3\msb@29

\mathchardef\nsubseteq="3\msb@2A

\mathchardef\nsupseteq="3\msb@2B

\mathchardef\nparallel="3\msb@2C

\mathchardef\nmid="3\msb@2D

\mathchardef\nshortmid="3\msb@2E

\mathchardef\nshortparallel="3\msb@2F

\mathchardef\nvdash="3\msb@30

\mathchardef\nVdash="3\msb@31

\mathchardef\nvDash="3\msb@32

\mathchardef\nVDash="3\msb@33

\mathchardef\ntrianglerighteq="3\msb@34

\mathchardef\ntrianglelefteq="3\msb@35

\mathchardef\ntriangleleft="3\msb@36

\mathchardef\ntriangleright="3\msb@37

\mathchardef\nleftarrow="3\msb@38

\mathchardef\nrightarrow="3\msb@39

\mathchardef\nLeftarrow="3\msb@3A

\mathchardef\nRightarrow="3\msb@3B

\mathchardef\nLeftrightarrow="3\msb@3C

\mathchardef\nleftrightarrow="3\msb@3D

\mathchardef\divideontimes="2\msb@3E

\mathchardef\varnothing="0\msb@3F

\mathchardef\nexists="0\msb@40

\mathchardef\mho="0\msb@66

\mathchardef\eth="0\msb@67

\mathchardef\eqsim="3\msb@68

\mathchardef\beth="0\msb@69

\mathchardef\gimel="0\msb@6A

\mathchardef\daleth="0\msb@6B

\mathchardef\lessdot="3\msb@6C

\mathchardef\gtrdot="3\msb@6D

\mathchardef\ltimes="2\msb@6E

\mathchardef\rtimes="2\msb@6F

\mathchardef\shortmid="3\msb@70

\mathchardef\shortparallel="3\msb@71

\mathchardef\smallsetminus="2\msb@72

\mathchardef\thicksim="3\msb@73

\mathchardef\thickapprox="3\msb@74

\mathchardef\approxeq="3\msb@75

\mathchardef\succapprox="3\msb@76

\mathchardef\precapprox="3\msb@77

\mathchardef\curvearrowleft="3\msb@78

\mathchardef\curvearrowright="3\msb@79

\mathchardef\digamma="0\msb@7A

\mathchardef\varkappa="0\msb@7B

\mathchardef\hslash="0\msb@7D

\mathchardef\hbar="0\msb@7E

\mathchardef\backepsilon="3\msb@7F

\def\Bbb{\ifmmode\let\next\Bbb@\else

 \def\next{\errmessage{Use \string\Bbb\space only in math mode}}\fi\next}

\def\Bbb@#1{{\Bbb@@{#1}}}

\def\Bbb@@#1{\fam\msbfam#1}

\catcode`\@=12